\documentclass[10pt]{article}
\pdfoutput=1

\usepackage[utf8]{inputenc}
\usepackage[T2A]{fontenc}

\input{epsf}
\usepackage{epsfig}
\usepackage{amssymb}
\usepackage{amsfonts}
\usepackage{amsbsy}
\usepackage[cmtip,all]{xy}
\usepackage{amsmath}
\usepackage{esint}
\usepackage{collectbox}
\usepackage{amscd}
\usepackage{mathrsfs}
\usepackage{amsmath,amsthm}
\usepackage{enumitem}
\usepackage{dsfont}
\usepackage{soul}
\usepackage{comment}
\usepackage{cite} 
\usepackage[most]{tcolorbox}
\usepackage{stmaryrd}
\usepackage{xcolor}
\definecolor{burgundy}{rgb}{0.5, 0.0, 0.13}
\definecolor{olive}{rgb}{0.50, 0.50, 0.0}
\usepackage[linktocpage=true,colorlinks=true,linkcolor=burgundy,citecolor=black!20!blue,urlcolor=violet]{hyperref}
\usepackage{accents}
\usepackage{xfrac}

\usepackage{tikz}
\usepackage{tikz-cd}
\usetikzlibrary{arrows}
\usetikzlibrary{arrows.meta}
\usetikzlibrary{positioning}
\usetikzlibrary{shapes,snakes}
\usetikzlibrary{fit}
\usetikzlibrary{decorations.pathmorphing,decorations.pathreplacing,decorations.markings}
\usetikzlibrary{calc}

\theoremstyle{definition}

\DeclareMathAlphabet{\mathpzc}{OT1}{pzc}{m}{it}

\def\exp{{\rm exp}}

\def\CA{{\cal A}}
\def\CB{{\cal B}}

\def\CB {{\cal B}}

\def\IC{\mathbb{C}}

\def\IZ{{\mathbb{Z}}}

\def\fs{\mathfrak{s}}

\def\fs{\mathfrak{s}}

\def\fu{\mathfrak{u}}

\def\lm{\limits}

\hoffset= - 1.0in

\numberwithin{equation}{section}

\tcbset{boxrule=0.6pt,colback=white!97!green}

\DeclareSymbolFont{bbsymbol}{U}{bbold}{m}{n}
\DeclareMathSymbol{\bbzero}{\mathbin}{bbsymbol}{"30}
\DeclareMathSymbol{\bbone}{\mathbin}{bbsymbol}{"31}
\DeclareMathSymbol{\bbtwo}{\mathbin}{bbsymbol}{"32}
\DeclareMathSymbol{\bbthree}{\mathbin}{bbsymbol}{"33}
\DeclareMathSymbol{\bbfour}{\mathbin}{bbsymbol}{"34}
\DeclareMathSymbol{\bbfive}{\mathbin}{bbsymbol}{"35}
\DeclareMathSymbol{\bbsix}{\mathbin}{bbsymbol}{"36}
\DeclareMathSymbol{\bbseven}{\mathbin}{bbsymbol}{"37}
\DeclareMathSymbol{\bbeight}{\mathbin}{bbsymbol}{"38}
\DeclareMathSymbol{\bbnine}{\mathbin}{bbsymbol}{"39}

\def\myblue{white!40!blue}

\newcommand\sqbox[1]{{
	\setbox0=\hbox{\mbox{$\Box$}}
	\setbox1=\hbox{\mbox{\raisebox{0.35ex}{\tiny #1}}}
	\mbox{\raisebox{-0.2ex}{\rlap{\hbox to \wd0{\hss{\box1}\hss}}\box0}}
}}

\begin{document}

\hfill MIPT/TH-20/24

\hfill ITEP/TH-26/24

\hfill IITP/TH-21/24

\vskip 1.5in
\begin{center}
	
	{\bf\Large On geometric bases for {\it quantum} A-polynomials of knots}
	
	\vskip 0.2in
	\renewcommand{\thefootnote}{\fnsymbol{footnote}}
	{Dmitry Galakhov$^{2,3,4,}$\footnote[2]{e-mail: galakhov@itep.ru} and  Alexei Morozov$^{1,2,3,4,}$\footnote[3]{e-mail: morozov@itep.ru}}
	\vskip 0.2in
	\renewcommand{\thefootnote}{\roman{footnote}}
	{\small{
			\textit{$^1$MIPT, 141701, Dolgoprudny, Russia}
			\vskip 0 cm
			\textit{$^2$NRC “Kurchatov Institute”, 123182, Moscow, Russia}
			\vskip 0 cm
			\textit{$^3$IITP RAS, 127051, Moscow, Russia}
			\vskip 0 cm
			\textit{$^4$ITEP, Moscow, Russia}
	}}
\end{center}

\vskip 0.2in
\baselineskip 16pt

\centerline{ABSTRACT}

\bigskip

{\footnotesize
	A simple geometric way is suggested to derive the Ward identities in the Chern-Simons theory,
	also known as quantum $A$- and $C$-polynomials for knots.
	In quasi-classical limit it is closely related to the well publicized augmentation theory and contact geometry.
	Quantization allows to present it in much simpler terms,
	what could make these techniques available to a broader audience.
	To avoid overloading of the presentation,
	only the case of the colored Jones polynomial for the trefoil knot is considered,
	though various generalizations are straightforward.
	Restriction to solely Jones polynomials (rather than full HOMFLY-PT)
	is related to a serious simplification, provided by the use  of Kauffman calculus.
	Going beyond looks realistic, however it remains a problem, both challenging and promising.
	
}

\bigskip

\bigskip

\tableofcontents

\bigskip

\section{Introduction}

The theory of knot polynomials is one of the simplest examples of non-perturbative
gauge theories which can be solved exactly --
thus providing us with unique experience and inspiration.
Conceptually it describes the Wilson line averages \cite{LL,Polyakov:1976fu,Polyakov:1987hqn}
in the Chern-Simons theory \cite{CS,Witten,Guad,MoSmi,Moore:1988qv}
as the functions of exponentials $q=\exp\frac{2\pi \sqrt{-1}}{g+N}$
where $g$ is the coupling constant and $N$ parameterizes the gauge group $SU(N)$.
Since the theory is topological, the averages depend only on the linking number of contours
and the resulting functions are polynomials (what is still a mystery from the point of view of QFT
-- whereas this observation is made technically obvious in, say, the Reshetikhin-Turaev formalism
\cite{RT1,RT2,1112.2654}).
Availability and abundance \cite{katlas}  of concrete answers for knot and link polynomials allows
examination of various ideas about the structure of QFT.

In particular, of interest is the subject of Ward identities.
In knot theory they acquire the form of quantum ${\cal A}$- and ${\cal C}$-polynomials \cite{GarA,GarC,2009.11641}.
In the simplest case these are the finite-difference equations w.r.t. the representation label $j$ of
symmetrically colored Jones polynomials $J_{[j]}(q)$ (i.e. HOMFLY polynomials for $N=2$),
which in the quasi-classical limit reduce to the classical $A$-polynomials \cite{cooper1994plane,cooper1998representation,Gukov:2003na,2019arXiv190301732D}.
Despite their conceptual significance and numerous applications \cite{Kashaev:1996kc, murakami2001colored, Gukov:2006ze, A=B, Arthamonov:2013rfa},
the quantum ${\cal A}$-polynomials
(not to say about the still less-popular, though, perhaps, more significant ${\cal C}$-ones)
remain exotics -- largely because of the lack of clear derivations or severe technicality of their presentations \cite{garoufalidis2005colored,Garoufalidis:2012rt}.
In this short note we outline a conceptually simple approach to the subject,
based on the transformation of links around the given knots.

The  point is that Ward identities are nothing but equivalencies between the topologically equal
(Reidemeister related) {\it link diagrams}.
At the same time for $N=2$ these diagrams can be ``planarized'' with the help of Kauffman's
$R$-matrix \cite{kauffman1987state,kauffman1989,kauffman1992link,Dolotin:2012sw} what reduces the corresponding link polynomials to linear combinations
of the simple basic elements (represented by links, which embrace the sets of strands in the knot).
In this way one expands in the same basis the Reidemeister equivalent diagrams --
and this allows to reduce the basis to just a few independent links.
Then expressing through the same basis the cabled knots
(i.e. the links, obtained by adding lines along the knot),
we obtain a linear relation between different additions -- what is exactly an equation between them
(between different cablings),
i.e. the quantum ${\cal A}$-polynomial.
This is the procedure, implemented implicitly in the contact geometry considerations in
{\it augmentation} theory \cite{2002math.....10124E,ekholm2013knot,Aganagic:2013jpa,Aganagic:2012jb,1210.4803}
-- but for us it looks too well-hidden in those wonderful papers.
As usual, the full-fledged quantum case is much simpler than the quasi-classical calculus.
Therefore we attempted to express the construction in the maximally straightforward terms
and comment on relation to augmentation theory at the very end.
In this note we do this for the simplest case of the trefoil -- to separate the idea from technicalities.
More examples will be presented elsewhere.

\section{Paradigm}

\subsection{Operators}

Let us consider a tubular neighborhood of a knot $K$ inside $S^3$.
Its tubular boundary is a topological torus $T^2$.
So we could consider $S^3\setminus K$ as a torus body.
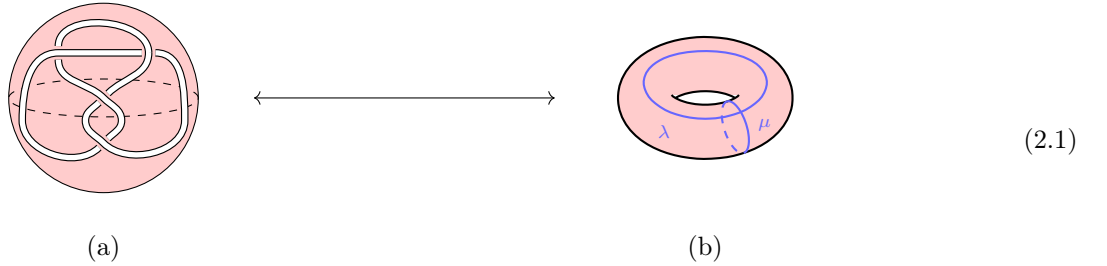
\begin{equation}\label{inversion}
	\begin{array}{c}
		\begin{tikzpicture}
			\begin{scope}[scale=0.6]
				\begin{scope}[shift={(0,1)}]
					\draw[fill=white!80!red] (0,-1) circle (2.1);
					\begin{scope}[shift={(0,-1)}]
						\begin{scope}[yscale=0.2]
							\draw[thin, dashed] (0,0) circle (2.1);
						\end{scope}
					\end{scope}
					\draw[line width=1mm] (1,0) to[out=90,in=90] (-1,0.15) (-1,-0.15) to[out=270,in=135] (0,-1) to[out=315,in=90] (0.4,-1.5) to[out=270,in=45] (0.1,-1.9) (-0.1,-2.1) to[out=225,in=270] (-1.8,-1.5) to[out=90,in=180] (-1,0) -- (0.85,0) (1.15,0) to[out=0,in=90] (1.8,-1.5) to[out=270, in=315] (0,-2) to[out=135,in=270] (-0.4,-1.5) to[out=90,in=225] (-0.1,-1.1) (0.1,-0.9) to[out=45,in=270] (1,0);
					\draw[line width=0.7mm,white] (1,0) to[out=90,in=90] (-1,0.15) (-1,-0.15) to[out=270,in=135] (0,-1) to[out=315,in=90] (0.4,-1.5) to[out=270,in=45] (0.1,-1.9) (-0.1,-2.1) to[out=225,in=270] (-1.8,-1.5) to[out=90,in=180] (-1,0) -- (0.85,0) (1.15,0) to[out=0,in=90] (1.8,-1.5) to[out=270, in=315] (0,-2) to[out=135,in=270] (-0.4,-1.5) to[out=90,in=225] (-0.1,-1.1) (0.1,-0.9) to[out=45,in=270] (1,0);
				\end{scope}
			\end{scope}
			\begin{scope}[shift={(8,0)}]
				\begin{scope}[scale=0.4]
					\draw[thick,fill=white!80!red] (-2.9,0.) to[out=-89.8399,in=100.017] (-2.87723,-0.26343) to[out=-79.9828,in=118.828] (-2.70107,-0.76265) to[out=-61.1725,in=134.258] (-2.40139,-1.16537) to[out=-45.742,in=146.399] (-2.02811,-1.47359) to[out=-33.6013,in=156.101] (-1.60625,-1.70413) to[out=-23.8987,in=164.195] (-1.14839,-1.86896) to[out=-15.8051,in=171.311] (-0.66453,-1.97387) to[out=-8.68872,in=184.42] (0.33791,-2.01131) to[out=4.41993,in=198.572] (1.30922,-1.81925) to[out=18.5724,in=207.161] (1.75584,-1.63275) to[out=27.1611,in=217.636] (2.16277,-1.37722) to[out=37.6355,in=230.875] (2.51464,-1.03837) to[out=50.8749,in=247.59] (2.7788,-0.60076) to[out=67.5902,in=269.84] (2.9,0.)
					(2.9,0.) to[out=90.1601,in=292.698] (2.7788,0.60076) to[out=112.698,in=309.358] (2.51464,1.03837) to[out=129.358,in=322.547] (2.16277,1.37722) to[out=142.547,in=332.985] (1.75584,1.63275) to[out=152.985,in=341.551] (1.30922,1.81925) to[out=161.551,in=355.683] (0.33791,2.01131) to[out=175.683,in=8.79496] (-0.66453,1.97387) to[out=-171.205,in=15.9226] (-1.14839,1.86896) to[out=-164.077,in=24.036] (-1.60625,1.70413) to[out=-155.964,in=33.7698] (-2.02811,1.47359) to[out=-146.23,in=45.956] (-2.40139,1.16537) to[out=-134.044,in=61.4422] (-2.70107,0.76265) to[out=-118.558,in=80.2966] (-2.87723,0.26343) to[out=-99.7034,in=89.8399] (-2.9,0.);
					\draw[thick,fill=white]
					(-1.11956,0.09908) to[out=-49.9491,in=153.423] (-0.92834,-0.03418) to[out=-26.5774,in=161.919] (-0.72701,-0.11594) to[out=-18.081,in=169.27] (-0.47349,-0.18059) to[out=-10.7297,in=175.99] (-0.18692,-0.21756) to[out=-4.01049,in=184.113] (0.18692,-0.21756) to[out=4.11276,in=190.839] (0.47349,-0.18059) to[out=10.8386,in=198.203] (0.72701,-0.11594) to[out=18.2034,in=206.723] (0.92834,-0.03418) to[out=26.7226,in=229.949] (1.11956,0.09908)
					(0.97323,0.01035) to[out=150.737,in=339.671] (0.78968,0.09405) to[out=159.671,in=347.286] (0.54899,0.16486) to[out=167.286,in=354.139] (0.2697,0.21034) to[out=174.139,in=357.416] (0.12181,0.22137) to[out=177.416,in=1.0626] (-0.04693,0.22375) to[out=-178.937,in=5.96464] (-0.2697,0.21034) to[out=-174.035,in=12.8258] (-0.54899,0.16486) to[out=-167.174,in=20.4567] (-0.78968,0.09405) to[out=-159.543,in=29.2627] (-0.97323,0.01035);
					\draw[thick,\myblue]
					(-2.,0.74437) to[out=-110.513,in=111.519] (-1.99235,0.25547) to[out=-68.4811,in=143.432] (-1.59517,-0.22917) to[out=-36.5678,in=164.608] (-0.84954,-0.57841) to[out=-15.3925,in=173.316] (-0.39091,-0.66766) to[out=-6.68404,in=182.646] (0.15119,-0.68718) to[out=2.64617,in=195.535] (0.84954,-0.57841) to[out=15.5348,in=216.766] (1.59517,-0.22917) to[out=36.7659,in=248.787] (1.99235,0.25547) to[out=68.7869,in=290.513] (2.,0.74437)
					(2.,0.74437) to[out=110.827,in=311.265] (1.85392,0.98337) to[out=131.265,in=335.001] (1.41484,1.29737) to[out=155.001,in=348.618] (0.84207,1.47882) to[out=168.618,in=357.467] (0.21479,1.55296) to[out=177.467,in=5.21746] (-0.42348,1.53888) to[out=-174.783,in=15.2103] (-1.03897,1.4325) to[out=-164.79,in=31.5016] (-1.57881,1.20973) to[out=-148.498,in=43.8516] (-1.79277,1.04724) to[out=-136.148,in=69.173] (-2.,0.74437);
					\node[\myblue, below left] at (-0.84954,-0.57841) {$\scriptstyle\lambda$};
					\draw[\myblue, thick]
					(1.27781,-1.82961) to[out=18.1969,in=232.273] (1.36305,-1.76682) to[out=52.2727,in=255.341] (1.42518,-1.63128) to[out=75.3413,in=270.783] (1.4499,-1.39568) to[out=90.783,in=277.359] (1.4373,-1.22373) to[out=97.3594,in=285.102] (1.38531,-0.96416) to[out=105.102,in=295.592] (1.25201,-0.60221) to[out=115.592,in=310.13] (1.05809,-0.29197) to[out=130.13,in=326.502] (0.91736,-0.16139) to[out=146.502,in=17.7616] (0.72219,-0.11751);
					\draw[\myblue, dashed, thick]
					(0.72219,-0.11751) to[out=-161.803,in=69.0245] (0.59176,-0.26297) to[out=-110.975,in=86.938] (0.55164,-0.47351) to[out=-93.062,in=100.849] (0.58046,-0.83433) to[out=-79.151,in=109.283] (0.66132,-1.13397) to[out=-70.717,in=115.835] (0.75151,-1.35224) to[out=-64.1654,in=122.344] (0.84463,-1.5209) to[out=-57.6557,in=135.66] (0.99829,-1.71626) to[out=-44.3405,in=148.487] (1.09512,-1.79371) to[out=-31.5133,in=197.762] (1.27781,-1.82961);
					\node[\myblue, above right] at (1.4499,-1.39568) {$\scriptstyle\mu$};
				\end{scope}
			\end{scope}
			\draw[<->] (2,0) -- (6,0);
			\node at (0,-2) {(a)};
			\node at (8,-2) {(b)};
		\end{tikzpicture}
	\end{array}
\end{equation}
CS path integral calculated over $S^3\setminus K$ with some choice of boundary conditions on $T^2$ provides a wave function $\Psi_K\in\mathscr{H}(T^2)$ \cite{Moore:1988qv,Elitzur:1989nr}.
Further specifying a boundary condition as a spin-$j$ Wilson line projects the wave function to a fixed polarization and a value of ``coordinate'' -- $\Psi_K(j)$.

In Hilbert space $\mathscr{H}(T^2)$ there are two natural operators corresponding to meridional $\mu$ and longitudinal $\lambda$ spin-$\sfrac{1}{2}$ Wilson lines.
Let us allow these operators to ``sink'' inside the torus body, then on the inverted side (\ref{inversion}a) they are translated to spin-$\sfrac{1}{2}$ Wilson strands linked to knot $K$.
In particular, operator $\mu$ represents a ring encircles $K$ anywhere (also we could have called it a Hopf link operator), whereas operator $\lambda$ converts the strand of $K$ into a cable consisting of two strands of spin-$j$ and spin-$\sfrac{1}{2}$ respectively:
\begin{equation}\label{operators}
	\mu\circ\begin{array}{c}
		\begin{tikzpicture}
			\draw[line width = 0.5mm] (0,-0.5) -- (0,0.5);
			\node[right] at (0,-0.5) {$\scriptstyle j$};
		\end{tikzpicture}
	\end{array}=\begin{array}{c}
		\begin{tikzpicture}
			\draw[line width = 0.5mm] (0,-0.5) -- (0,-0.3) (0,-0.2) -- (0,0.5);
			\draw[\myblue,line width = 0.5mm] (-0.1,0.25) to[out=180,in=90] (-0.5,0) to[out=270,in=180] (0,-0.25) to[out=0,in=270] (0.5,0) to[out=90,in=0] (0.1,0.25);
			\node[right] at (0,-0.5) {$\scriptstyle j$};
		\end{tikzpicture}
	\end{array},\quad 	\lambda\circ\begin{array}{c}
		\begin{tikzpicture}
			\draw[line width = 0.5mm] (0,-0.5) -- (0,0.5);
			\node[right] at (0,-0.5) {$\scriptstyle j$};
		\end{tikzpicture}
	\end{array}=\begin{array}{c}
		\begin{tikzpicture}
			\draw[line width = 0.5mm] (0,-0.5) -- (0,0.5);
			\draw[\myblue,line width = 0.5mm] (-0.3,-0.5) -- (-0.3,0.5);
			\node[right] at (0,-0.5) {$\scriptstyle j$};
		\end{tikzpicture}
	\end{array}\,.
\end{equation}
Here and in what follows we will denote the spin-$\sfrac{1}{2}$ strand by blue color.

It is straightforward to calculate the action of these operators on the wave function:
\begin{equation}
	\mu\,\Psi_K(j)=-\left(q^{2j+1}+q^{-2j-1}\right)\,\Psi_K(j),\quad\lambda\,\Psi_K(j)=\Psi_K\left(j+\frac{1}{2}\right)+\Psi_K\left(j-\frac{1}{2}\right)\,.
\end{equation}

\subsection{Link symbols}

As we have seen operators acting on the wave-functions can be represented as certain links $L$ entangled with $K$.
For a link $L$ let us introduce a symbol $L$ acting on the wave function:
\begin{equation}
	L\Psi_K(j):=\Psi_{L K}(j)\,.
\end{equation}

{\bf\color{burgundy}Warning:} a link symbol is not an operator (does not belong to an algebra of operators on $\mathscr{H}(T^2)$) in general.
As a generic link can not be all pushed to a small tubular neighborhood of $K$.
In particular, as we will see in what follows, a multiplication of link symbols as linear operators does not work!

\begin{tcolorbox}
	{\bf Conjecture:} a vector space of link symbols $(L_1,L_2,L_3,\ldots)$ over field $\IZ[\mu,q,q^{-1}]$ has a finite basis.
\end{tcolorbox}

Let us suppose the dimension of this basis is $d$.
Consider a set of vectors in this link space:
\begin{equation}
	\lambda^n\,\Psi_K(j)=\sum\lm_{i=1}^d c_{ni}(\mu)L_i\Psi_{K}(j)\,.
\end{equation}
If we consider $n=1,\ldots,d+1$ then these are $d+1$ vector in a $d$-dimensional vector space.
They can not be linearly independent.
Then there exists a \emph{quantum} $\cal A$-polynomial imposing a relation on these vectors:
\begin{tcolorbox}
	\begin{equation}
		\CA_K(\mu,\lambda)\,\Psi_K(j)=\sum\lm_{n=1}^{d+1}a_n(\mu)\,\lambda^n\,\Psi_K(j)=0\,.
	\end{equation}
\end{tcolorbox}

\subsection{Link planarization}

At the moment we are not aware of a generic method to produce a basis in the link symbol space.
As a disclaimer comment let us emphasize that this problem is notoriously difficult and rather ubiquitous.
In general, to produce Ward identities for some partition function we calculate derivatives with respect to some external currents $y$, fugacities etc.
These derivatives bring down expectation values of $\delta_y S$, where $S$ is some action.
On the other hand the usual trick of varying integration variables $x$ in the path integral tells us that the ring of operators is defined modulo e.o.m. $\delta_x S$.
Finding any Gr\"obner basis in such a ring and then constructing an overdetermined system of operators  $\delta_y S$ proves to be rather difficult unless some certain appropriate intrinsic symmetries of the system in question are involved.
Surely, some higher form symmetry would be rather helpful, yet in an indirect way, as $\delta_x S$ are usually local operators.
As disappearance of symmetry makes a ``difficulty'' of Ward identities to jump abruptly one might follow already in a canonical example of Feynman graphs \cite{Mishnyakov:2024rmb}.
As we will see in what follows in our example of A-polynomials a somewhat simplifying, yet loosing important information way is a transition to quasi-classics where quantum operators (as well as our link symbols) turn into commutative variables, and one might try to apply the Galois theory for fields of the latters.

Nevertheless a rather effective method to produce linear relations between link symbols is a link planarization, or a reduction procedure, consisting of two stages: first we flatten a link, then we apply the Kauffman bracket to eliminate appearing self-intersections.

Flattening procedure implies simply that the whole link may be transported smoothly to a single temporal slice in the knot evolution diagram representation.
Under this transport the link strands are acted upon by the braid group:
\begin{equation}\label{braid}
	\begin{array}{c}
		\begin{tikzpicture}
			\draw[line width = 0.5mm,\myblue] (0.2,1) -- (0.2,-0.5) to[out=270,in=0] (0,-0.7);
			\draw[line width = 2mm,white] (-0.5,1) -- (-0.5,0.5) to[out=270,in=90] (0.5,-0.5) -- (0.5,-1);
			\draw[line width = 1.2mm,black] (-0.5,1) -- (-0.5,0.5) to[out=270,in=90] (0.5,-0.5) -- (0.5,-1);
			\draw[line width = 1mm,orange] (-0.5,1) -- (-0.5,0.5) to[out=270,in=90] (0.5,-0.5) -- (0.5,-1);
			\draw[line width = 2mm,white] (0.5,1) -- (0.5,0.5) to[out=270,in=90] (-0.5,-0.5) -- (-0.5,-1);
			\draw[line width = 1.2mm,black] (0.5,1) -- (0.5,0.5) to[out=270,in=90] (-0.5,-0.5) -- (-0.5,-1);
			\draw[line width = 1mm,black!60!green] (0.5,1) -- (0.5,0.5) to[out=270,in=90] (-0.5,-0.5) -- (-0.5,-1);
			\draw[line width = 1.5mm,white] (-0.2,1) -- (-0.2,-0.5) to[out=270,in=180] (0,-0.7);
			\draw[line width = 0.5mm,\myblue] (-0.2,1) -- (-0.2,-0.5) to[out=270,in=180] (0,-0.7);
			\begin{scope}[shift={(-1,0)}]
				\draw[-stealth] (0,-1) -- (0,1);
				\node[left] at (0,1) {$\scriptstyle t$};
			\end{scope}
		\end{tikzpicture}\\
		\hline
		t=0:\,\begin{array}{c}
			\begin{tikzpicture}
				\draw[fill=black!60!green] (-0.5,0) circle (0.1);
				\draw[fill=orange] (0.5,0) circle (0.1);
				\draw[line width = 0.5mm,\myblue] (0,-0.5) -- (0,0.5);
			\end{tikzpicture}
		\end{array}
	\end{array}\;\overset{R}{\longrightarrow}\;\begin{array}{c}
		\begin{tikzpicture}
			\draw[line width = 0.5mm,\myblue] (0.2,1) to[out=270,in=90] (-0.7,0.8);
			\draw[line width = 2mm,white] (-0.5,1) -- (-0.5,0.5) to[out=270,in=90] (0.5,-0.5) -- (0.5,-1);
			\draw[line width = 1.2mm,black] (-0.5,1) -- (-0.5,0.5) to[out=270,in=90] (0.5,-0.5) -- (0.5,-1);
			\draw[line width = 1mm,orange] (-0.5,1) -- (-0.5,0.5) to[out=270,in=90] (0.5,-0.5) -- (0.5,-1);
			\draw[line width = 1.5mm,white]  (-0.7,0.8) to[out=270,in=90] (0.7,0.6);
			\draw[line width = 0.5mm,\myblue]  (-0.7,0.8) to[out=270,in=90] (0.7,0.6);
			\draw[line width = 2mm,white] (0.5,1) -- (0.5,0.5) to[out=270,in=90] (-0.5,-0.5) -- (-0.5,-1);
			\draw[line width = 1.2mm,black] (0.5,1) -- (0.5,0.5) to[out=270,in=90] (-0.5,-0.5) -- (-0.5,-1);
			\draw[line width = 1mm,black!60!green] (0.5,1) -- (0.5,0.5) to[out=270,in=90] (-0.5,-0.5) -- (-0.5,-1);
			\draw[line width = 1.5mm,white]  (0.7,0.6) to[out=270,in=270] (-0.2,1);
			\draw[line width = 0.5mm,\myblue]  (0.7,0.6) to[out=270,in=270] (-0.2,1);
			\begin{scope}[shift={(-1,0)}]
				\draw[-stealth] (0,-1) -- (0,1);
				\node[left] at (0,1) {$\scriptstyle t$};
			\end{scope}
		\end{tikzpicture}\\
		\hline
		t=1:\,\begin{array}{c}
			\begin{tikzpicture}
				\draw[fill=orange] (-0.5,0) circle (0.1);
				\draw[fill=black!60!green] (0.5,0) circle (0.1);
				\draw[line width = 0.5mm,\myblue] (0,-0.5) to[out=90,in=270] (0.7,0) to[out=90,in=0] (0.5,0.3) to[out=180,in=0] (-0.5,-0.3) to[out=180,in=270] (-0.7,0) to[out=90,in=270] (0,0.5);
			\end{tikzpicture}
		\end{array}
	\end{array},\quad
	\begin{array}{c}
		\begin{tikzpicture}[xscale=-1]
			\draw[line width = 0.5mm,\myblue] (0.2,1) -- (0.2,-0.5) to[out=270,in=0] (0,-0.7);
			\draw[line width = 2mm,white] (-0.5,1) -- (-0.5,0.5) to[out=270,in=90] (0.5,-0.5) -- (0.5,-1);
			\draw[line width = 1.2mm,black] (-0.5,1) -- (-0.5,0.5) to[out=270,in=90] (0.5,-0.5) -- (0.5,-1);
			\draw[line width = 1.2mm,black!60!green] (-0.5,1) -- (-0.5,0.5) to[out=270,in=90] (0.5,-0.5) -- (0.5,-1);
			\draw[line width = 2mm,white] (0.5,1) -- (0.5,0.5) to[out=270,in=90] (-0.5,-0.5) -- (-0.5,-1);
			\draw[line width = 1.2mm,black] (0.5,1) -- (0.5,0.5) to[out=270,in=90] (-0.5,-0.5) -- (-0.5,-1);
			\draw[line width = 1mm,orange] (0.5,1) -- (0.5,0.5) to[out=270,in=90] (-0.5,-0.5) -- (-0.5,-1);
			\draw[line width = 1.5mm,white] (-0.2,1) -- (-0.2,-0.5) to[out=270,in=180] (0,-0.7);
			\draw[line width = 0.5mm,\myblue] (-0.2,1) -- (-0.2,-0.5) to[out=270,in=180] (0,-0.7);
			\begin{scope}[shift={(1,0)}]
				\draw[-stealth] (0,-1) -- (0,1);
				\node[left] at (0,1) {$\scriptstyle t$};
			\end{scope}
		\end{tikzpicture}\\
		\hline
		t=0:\,\begin{array}{c}
			\begin{tikzpicture}
				\draw[fill=black!60!green] (-0.5,0) circle (0.1);
				\draw[fill=orange] (0.5,0) circle (0.1);
				\draw[line width = 0.5mm,\myblue] (0,-0.5) -- (0,0.5);
			\end{tikzpicture}
		\end{array}
	\end{array}\;\overset{R^{-1}}{\longrightarrow}\;\begin{array}{c}
		\begin{tikzpicture}[xscale=-1]
			\draw[line width = 0.5mm,\myblue] (0.2,1) to[out=270,in=90] (-0.7,0.8);
			\draw[line width = 2mm,white] (-0.5,1) -- (-0.5,0.5) to[out=270,in=90] (0.5,-0.5) -- (0.5,-1);
			\draw[line width = 1.2mm,black] (-0.5,1) -- (-0.5,0.5) to[out=270,in=90] (0.5,-0.5) -- (0.5,-1);
			\draw[line width = 1mm,black!60!green] (-0.5,1) -- (-0.5,0.5) to[out=270,in=90] (0.5,-0.5) -- (0.5,-1);
			\draw[line width = 1.5mm,white]  (-0.7,0.8) to[out=270,in=90] (0.7,0.6);
			\draw[line width = 0.5mm,\myblue]  (-0.7,0.8) to[out=270,in=90] (0.7,0.6);
			\draw[line width = 2mm,white] (0.5,1) -- (0.5,0.5) to[out=270,in=90] (-0.5,-0.5) -- (-0.5,-1);
			\draw[line width = 1.2mm,black] (0.5,1) -- (0.5,0.5) to[out=270,in=90] (-0.5,-0.5) -- (-0.5,-1);
			\draw[line width = 1mm,orange] (0.5,1) -- (0.5,0.5) to[out=270,in=90] (-0.5,-0.5) -- (-0.5,-1);
			\draw[line width = 1.5mm,white]  (0.7,0.6) to[out=270,in=270] (-0.2,1);
			\draw[line width = 0.5mm,\myblue]  (0.7,0.6) to[out=270,in=270] (-0.2,1);
			\begin{scope}[shift={(1,0)}]
				\draw[-stealth] (0,-1) -- (0,1);
				\node[left] at (0,1) {$\scriptstyle t$};
			\end{scope}
		\end{tikzpicture}\\
		\hline
		t=1:\,\begin{array}{c}
			\begin{tikzpicture}[xscale=-1]
				\draw[fill=black!60!green] (-0.5,0) circle (0.1);
				\draw[fill=orange] (0.5,0) circle (0.1);
				\draw[line width = 0.5mm,\myblue] (0,-0.5) to[out=90,in=270] (0.7,0) to[out=90,in=0] (0.5,0.3) to[out=180,in=0] (-0.5,-0.3) to[out=180,in=270] (-0.7,0) to[out=90,in=270] (0,0.5);
			\end{tikzpicture}
		\end{array}
	\end{array}\,.
\end{equation}

We present the flattened link diagrams as knot link diagrams for spin-$\sfrac{1}{2}$ strands in a plane with punctures representing strands of the very knot in question.
The resulting diagram appears as we move the point of view on the whole link diagram upwards, so that the observer looks at the link from above.

On the plane all the self-intersections of the spin-$\sfrac{1}{2}$ strand may be expanded over non-intersecting links with the help of the Kauffman rule \cite{kauffman1987state}:
\begin{equation}\label{Kauffman}
	\begin{aligned}
		&\begin{array}{c}
			\begin{tikzpicture}[scale=0.7]
				\draw[\myblue,line width = 0.5mm] (-0.5,-0.5) -- (0.5,0.5) (0.5,-0.5) -- (0.1,-0.1) (-0.1,0.1) --  (-0.5,0.5);
			\end{tikzpicture}
		\end{array}=q^{-\frac{1}{2}}\begin{array}{c}
			\begin{tikzpicture}[scale=0.7]
				\draw[\myblue,line width = 0.5mm] (-0.5,-0.5) to[out=45,in=315] (-0.5,0.5) (0.5,-0.5) to[out=135,in=225] (0.5,0.5);
			\end{tikzpicture}
		\end{array}\; + \; q^{\frac{1}{2}}\begin{array}{c}
			\begin{tikzpicture}[scale=0.7,rotate=90]
				\draw[\myblue,line width = 0.5mm] (-0.5,-0.5) to[out=45,in=315] (-0.5,0.5) (0.5,-0.5) to[out=135,in=225] (0.5,0.5);
			\end{tikzpicture}
		\end{array}\,,\\
		&\begin{array}{c}
			\begin{tikzpicture}[scale=0.7,rotate=90]
				\draw[\myblue,line width = 0.5mm] (-0.5,-0.5) -- (0.5,0.5) (0.5,-0.5) -- (0.1,-0.1) (-0.1,0.1) --  (-0.5,0.5);
			\end{tikzpicture}
		\end{array}=q^{\frac{1}{2}}\begin{array}{c}
			\begin{tikzpicture}[scale=0.7]
				\draw[\myblue,line width = 0.5mm] (-0.5,-0.5) to[out=45,in=315] (-0.5,0.5) (0.5,-0.5) to[out=135,in=225] (0.5,0.5);
			\end{tikzpicture}
		\end{array} \; + \; q^{-\frac{1}{2}}\begin{array}{c}
			\begin{tikzpicture}[scale=0.7,rotate=90]
				\draw[\myblue,line width = 0.5mm] (-0.5,-0.5) to[out=45,in=315] (-0.5,0.5) (0.5,-0.5) to[out=135,in=225] (0.5,0.5);
			\end{tikzpicture}
		\end{array}\,.
	\end{aligned}
\end{equation}

Thus among the planarized links we could choose a basis consisting of non-intersecting circles enveloping various punctures knot strands pierce in the fixed time slice.
Furthermore in the presence of a knot Reidemeister equivalences for these basis element together with Kauffman expansions of emerging intersections produce new relations among elements of the aforementioned basis.
Eventually, we conjectured that after Gaussian elimination the final basis is finite.

\section{Example: trefoil}

\subsection{Link symbols}

For a trefoil the link symbol basis consists of the following natural links for $p\in \IZ_{\geq 0}$:
\begin{equation}\label{Lp}
	L_p=\begin{array}{c}
		\begin{tikzpicture}[scale=0.7]
			\begin{scope}[shift={(0,-0.25)}]
				\begin{scope}[yscale=0.8]
					\draw[\myblue,line width = 0.5mm] (-0.7,0) to[out=90,in=90] (0.7,0);
				\end{scope}
			\end{scope}
			\foreach \y in {0,1,2}
			{
				\begin{scope}[shift={(0,\y)}]
					\draw[line width = 1.5mm,white] (0.5,0) to[out=90,in=270] (-0.5,1);
					\draw[line width = 0.5mm,black] (0.5,0) to[out=90,in=270] (-0.5,1);
					\draw[line width = 1.5mm,white] (-0.5,0) to[out=90,in=270] (0.5,1);
					\draw[line width = 0.5mm,black] (-0.5,0) to[out=90,in=270] (0.5,1);
				\end{scope}
			}
			\draw[line width = 1.5mm,white] (-0.5,0) -- (-0.5,-0.5) (0.5,0) -- (0.5,-0.5);
			\draw[line width = 0.5mm,black] (-0.5,0) -- (-0.5,-0.5) (0.5,0) -- (0.5,-0.5);
			\draw[line width = 0.5mm,black] (-0.5,-0.5) to[out=270,in=270] (-1,-0.5) -- (-1,3) to[out=90,in=90] (-0.5,3);
			\draw[line width = 0.5mm,black] (0.5,-0.5) to[out=270,in=270] (1,-0.5) -- (1,3) to[out=90,in=90] (0.5,3);
			\begin{scope}[shift={(0,-0.25)}]
				\begin{scope}[yscale=0.8]
					\draw[line width = 1.5mm,white] (-0.7,0) to[out=270,in=270] (0.7,0);
					\draw[\myblue,line width = 0.5mm] (-0.7,0) to[out=270,in=270] (0.7,0);
				\end{scope}
			\end{scope}
			\node[below] at (0,-0.7) {$\scriptstyle p$};
		\end{tikzpicture}
	\end{array}\,.
\end{equation}
where we put $p$ rings encircling two strands.

In what follows we will argue that the actual basis in the link symbol space for this trefoil representation is 2d, spanned by $L_0\Psi_{3_1}(j)=\Psi_{3_1}(j)$ and $L_1\Psi_{3_1}(j)$.

Let us consider the following link and reduce it in two ways -- flattening the side ring upwards and downwards:
\begin{equation}\label{two_paths}
	\begin{array}{c}
		\begin{tikzpicture}
			\node(A) at (0,0) {$\begin{array}{c}
					\begin{tikzpicture}[scale=1]
						\draw[-stealth] (-1.5,-0.2) -- (-1.5,3.7);
						\node[left] at (-1.5,3.7) {$\scriptstyle t$};
						\draw[ultra thick] (-1.55,0) -- (-1.45,0) (-1.55,1) -- (-1.45,1) (-1.55,2) -- (-1.45,2) (-1.55,3) -- (-1.45,3);
						\node[left] at (-1.5,0) {$0$};
						\node[left] at (-1.5,1) {$\sfrac{1}{3}$};
						\node[left] at (-1.5,2) {$\sfrac{2}{3}$};
						\node[left] at (-1.5,3) {$1$};
						\draw[\myblue,line width = 0.5mm] (1,3.5) to[out=180,in=90] (0.8,2) to[out=270,in=90] (0.3,1);
						\foreach \y in {0,1,2}
						{
							\begin{scope}[shift={(0,\y)}]
								\draw[line width = 1.5mm,white] (0.5,0) to[out=90,in=270] (-0.5,1);
								\draw[line width = 0.5mm,black] (0.5,0) to[out=90,in=270] (-0.5,1);
								\draw[line width = 1.5mm,white] (-0.5,0) to[out=90,in=270] (0.5,1);
								\draw[line width = 0.5mm,black] (-0.5,0) to[out=90,in=270] (0.5,1);
							\end{scope}
						}
						\draw[line width = 1.5mm,white] (-0.5,0) -- (-0.5,-0.5) (0.5,0) -- (0.5,-0.5);
						\draw[line width = 0.5mm,black] (-0.5,0) to[out=270,in=270] (-1,0) -- (-1,3) to[out=90,in=90] (-0.5,3);
						\draw[line width = 1.5mm,white] (0.5,0) to[out=270,in=270] (1.5,0) -- (1.5,3) to[out=90,in=90] (0.5,3);
						\draw[line width = 0.5mm,black] (0.5,0) to[out=270,in=270] (1.5,0) -- (1.5,3) to[out=90,in=90] (0.5,3);
						\draw[line width = 1.5mm,white] (0.3,1) to[out=270,in=270] (1.2,2) to[out=90,in=0] (1,3.5);
						\draw[\myblue,line width = 0.5mm] (0.3,1) to[out=270,in=270] (1.2,2) to[out=90,in=0] (1,3.5);
						\draw[thin, dashed, -stealth] (0.3,1) to[out=90,in=270] (-0.7,2) to[out=90,in=270] node[ pos=0.3,above left] {\tiny A} (0.3,3);
						\draw[thin, dashed, -stealth] (0.7,1) to[out=270,in=90] node[pos=0.2,right] {\tiny B} (-0.3,0);
						\draw[thin, dashed, -stealth] (1,3.5) to[out=0,in=90] (1.6,1.5) to[out=270,in=0] (1.2,-0.4) to[out=180,in=270] (0.4,0);
						\node[right] at (1.6,1.5) {\tiny B};
					\end{tikzpicture}
				\end{array}$};
			\node (B) at (5,1.3) {\scalebox{0.7}{$\begin{array}{c}
						\begin{tikzpicture}
							\draw[black,line width = 0.5mm] (-0.5,-0.5) -- (-0.5,0) (0.5,-0.5) -- (0.5,0);
							\begin{scope}[scale=0.5]
								\draw[white,line width = 1.5mm] (0,0.8) to[out=180,in=90] (-1.4,0) to[out=270,in=180] (-1,-0.4) to[out=0,in=180] (1,0.8) to[out=0,in=90] (1.8,0) to[out=270,in=0] (0,-0.8) to[out=180,in=270] (-1.8,0) to[out=90,in=180] (-1,0.8);
								\draw[\myblue,line width = 0.5mm] (0,0.8) to[out=180,in=90] (-1.4,0) to[out=270,in=180] (-1,-0.4) to[out=0,in=180] (1,0.8) to[out=0,in=90] (1.8,0) to[out=270,in=0] (0,-0.8) to[out=180,in=270] (-1.8,0) to[out=90,in=180] (-1,0.8);
								\draw[white,line width = 1.5mm] (-1,0.8) to[out=0,in=180] (1,-0.4) to[out=0,in=270] (1.4,0) to[out=90,in=0] (0,0.8);
								\draw[\myblue,line width = 0.5mm] (-1,0.8) to[out=0,in=180] (1,-0.4) to[out=0,in=270] (1.4,0) to[out=90,in=0] (0,0.8);
							\end{scope}
							\draw[white,line width = 1.5mm] (-0.5,0.5) -- (-0.5,0) (0.5,0.5) -- (0.5,0);
							\draw[black,line width = 0.5mm] (-0.5,0.5) -- (-0.5,0) (0.5,0.5) -- (0.5,0);
							\foreach \y in {0,-1,-2}
							{
								\begin{scope}[shift={(0,-0.75+0.5*\y)}]
									\begin{scope}[yscale=0.5]
										\draw[black,line width = 0.5mm] (0.5,-0.5) to[out=90,in=270] (-0.5,0.5);
										\draw[white,line width = 1.5mm] (-0.5,-0.5) to[out=90,in=270] (0.5,0.5);
										\draw[black,line width = 0.5mm] (-0.5,-0.5) to[out=90,in=270] (0.5,0.5);
									\end{scope}
								\end{scope}
							}
							\draw[black,line width = 0.5mm] (0.5,0.5) to[out=90,in=90] (1.0,0.5) -- (1.0,-2) to[out=270,in=270] (0.5,-2);
							\draw[black,line width = 0.5mm] (-0.5,0.5) to[out=90,in=90] (-1.0,0.5) -- (-1.0,-2) to[out=270,in=270] (-0.5,-2);
						\end{tikzpicture}
					\end{array}$}};
			\node (C) at (5,-1.3) {\scalebox{0.7}{$\begin{array}{c}
						\begin{tikzpicture}[scale=-1]
							\draw[black,line width = 0.5mm] (-0.5,0.5) -- (-0.5,0) (0.5,0.5) -- (0.5,0);
							\begin{scope}[scale=0.5]
								\draw[white,line width = 1.5mm] (-1,-0.4) to[out=0,in=180] (1,0.4);
								\draw[\myblue,line width = 0.5mm] (-1,-0.4) to[out=0,in=180] (1,0.4);
								\draw[white,line width = 1.5mm] (-1,0.4) to[out=0,in=180] (1,-0.4);
								\draw[\myblue,line width = 0.5mm] (-1,0.4) to[out=0,in=180] (1,-0.4);
								\draw[white,line width = 1.5mm] (-1,0.4) to[out=180,in=90] (-1.4,0) to[out=270,in=180] (-1,-0.4);
								\draw[\myblue,line width = 0.5mm] (-1,0.4) to[out=180,in=90] (-1.4,0) to[out=270,in=180] (-1,-0.4);
								\begin{scope}[xscale=-1]
									\draw[white,line width = 1.5mm] (-1,0.4) to[out=180,in=90] (-1.4,0) to[out=270,in=180] (-1,-0.4);
									\draw[\myblue,line width = 0.5mm] (-1,0.4) to[out=180,in=90] (-1.4,0) to[out=270,in=180] (-1,-0.4);
								\end{scope}
							\end{scope}
							\draw[white,line width = 1.5mm] (-0.5,-0.5) -- (-0.5,0) (0.5,-0.5) -- (0.5,0);
							\draw[black,line width = 0.5mm] (-0.5,-0.5) -- (-0.5,0) (0.5,-0.5) -- (0.5,0);
							\foreach \y in {0,-1,-2}
							{
								\begin{scope}[shift={(0,-0.75+0.5*\y)}]
									\begin{scope}[yscale=0.5]
										\draw[black,line width = 0.5mm] (0.5,-0.5) to[out=90,in=270] (-0.5,0.5);
										\draw[white,line width = 1.5mm] (-0.5,-0.5) to[out=90,in=270] (0.5,0.5);
										\draw[black,line width = 0.5mm] (-0.5,-0.5) to[out=90,in=270] (0.5,0.5);
									\end{scope}
								\end{scope}
							}
							\draw[black,line width = 0.5mm] (0.5,0.5) to[out=90,in=90] (1.0,0.5) -- (1.0,-2) to[out=270,in=270] (0.5,-2);
							\draw[black,line width = 0.5mm] (-0.5,0.5) to[out=90,in=90] (-1.0,0.5) -- (-1.0,-2) to[out=270,in=270] (-0.5,-2);
						\end{tikzpicture}
					\end{array}$}};
			\node(D) at (8,-1.3) {$\begin{array}{c}
					\begin{tikzpicture}[scale=0.7]
						\draw[fill=white] (-1,0) circle (0.1) (1,0) circle (0.1);
						\draw[\myblue,line width = 0.5mm] (-1,-0.4) to[out=0,in=180] (1,0.4);
						\draw[white,line width = 1.5mm] (-1,0.4) to[out=0,in=180] (1,-0.4);
						\draw[\myblue,line width = 0.5mm] (-1,0.4) to[out=0,in=180] (1,-0.4);
						\draw[\myblue,line width = 0.5mm] (-1,0.4) to[out=180,in=90] (-1.4,0) to[out=270,in=180] (-1,-0.4);
						\begin{scope}[xscale=-1]
							\draw[\myblue,line width = 0.5mm] (-1,0.4) to[out=180,in=90] (-1.4,0) to[out=270,in=180] (-1,-0.4);
						\end{scope}
					\end{tikzpicture}
				\end{array}$};
			\node(E) at (8,1.3) {$\begin{array}{c}
					\begin{tikzpicture}[scale=0.7]
						\draw[fill=white] (-1,0) circle (0.1) (1,0) circle (0.1);
						\draw[\myblue,line width = 0.5mm] (0,0.8) to[out=180,in=90] (-1.4,0) to[out=270,in=180] (-1,-0.4) to[out=0,in=180] (1,0.8) to[out=0,in=90] (1.8,0) to[out=270,in=0] (0,-0.8) to[out=180,in=270] (-1.8,0) to[out=90,in=180] (-1,0.8);
						\draw[white,line width = 1.5mm] (-1,0.8) to[out=0,in=180] (1,-0.4) to[out=0,in=270] (1.4,0) to[out=90,in=0] (0,0.8);
						\draw[\myblue,line width = 0.5mm] (-1,0.8) to[out=0,in=180] (1,-0.4) to[out=0,in=270] (1.4,0) to[out=90,in=0] (0,0.8);
					\end{tikzpicture}
				\end{array}$};
			\path (A) edge[double equal sign distance] node[sloped,above] {up (A)} (B) (A) edge[double equal sign distance] node[sloped,below] {down (B)} (C) (B) edge[double equal sign distance] (E) (C) edge[double equal sign distance] (D) (D) edge[double equal sign distance] (E);
			\node[right] at (9.5,-1.3) {at $t=0$};
			\node[right] at (9.5,1.3) {at $t=1$};
		\end{tikzpicture}
	\end{array}\,.
\end{equation}

To pass the lower half of the ring link through knot diagram intersections we applied braid rules \eqref{braid}:
\begin{equation}
	\begin{array}{c}
		\begin{tikzpicture}
			\begin{scope}[shift = {(0,0)}]
				\begin{scope}[scale=0.5]
					\draw (-1, 0) circle (0.1) (1, 0) circle (0.1);
					\draw[thick, \myblue] (2,0.5) to[out=180,in=0] (1,-0.3) to[out=180,in=0] (-1,0.3) to[out=180,in=90] (-1.3,0) to[out=270,in=180] (-1,-0.4) to[out=0,in=180] (2,-0.5);
				\end{scope}
			\end{scope}
			\begin{scope}[shift = {(3,0)}]
			\begin{scope}[scale=0.5]
				\draw (-1, 0) circle (0.1) (1, 0) circle (0.1);
				\draw[thick, \myblue] (2,0.5) to[out=180,in=90] (0,0) to[out=270,in=180] (2,-0.5);
			\end{scope}
			\end{scope}
			\begin{scope}[shift = {(6,0)}]
				\begin{scope}[scale=0.5,yscale=-1]
					\draw (-1, 0) circle (0.1) (1, 0) circle (0.1);
					\draw[thick, \myblue] (2,0.5) to[out=180,in=0] (1,-0.3) to[out=180,in=0] (-1,0.3) to[out=180,in=90] (-1.3,0) to[out=270,in=180] (-1,-0.4) to[out=0,in=180] (2,-0.5);
				\end{scope}
			\end{scope}
			\begin{scope}[shift = {(9,0)}]
				\begin{scope}[scale=0.5]
					\draw (-1, 0) circle (0.1) (1, 0) circle (0.1);
					\draw[thick, \myblue] (2,0.5) to[out=180,in=0] (1,0.7) to[out=180,in=0] (-1,0.5) to[out=180,in=90] (-1.5,0) to[out=270,in=180] (-1,-0.5) to[out=0,in=180] (1,-0.3) to[out=0,in=270] (1.3,0) to[out=90,in=0] (1,0.3) to[out=180,in=0] (-1,-0.3) to[out=180,in=270] (-1.3,0) to[out=90,in=180] (-1,0.3) to[out=0,in=180] (1,0.5) to[out=0,in=180] (2,-0.5);
				\end{scope}
			\end{scope}
			\begin{scope}[shift={(0,-0.7)}]
				\draw[-stealth] (-0.5,0) -- (9.5,0);
				\node[right] at (9.5,0) {$\scriptstyle t$};
				\draw[ultra thick] (0,-0.05) -- (0,0.05) (3,-0.05) -- (3,0.05) (6,-0.05) -- (6,0.05) (9,-0.05) -- (9,0.05);
				\node[below] at (0,-0.5) {$0$};
				\node[below] at (3,-0.5) {$\sfrac{1}{3}$};
				\node[below] at (6,-0.5) {$\sfrac{2}{3}$};
				\node[below] at (9,-0.5) {$1$};
			\end{scope}
			\draw[-stealth] (1.8,0) -- (1.2,0) node[pos=0.5,above] {$\scriptstyle R^{-1}$};
			\draw[-stealth] (4.2,0) -- (4.8,0) node[pos=0.5,above] {$\scriptstyle R$};
			\draw[-stealth] (7.2,0) -- (7.8,0) node[pos=0.5,above] {$\scriptstyle R$};
		\end{tikzpicture}
	\end{array}
\end{equation}

As a result we derive an equivalence of the following flattened links:
\begin{equation}
	\begin{array}{c}
		\begin{tikzpicture}[scale=0.7]
			\draw[fill=white] (-1,0) circle (0.1) (1,0) circle (0.1);
			\draw[\myblue,line width = 0.5mm] (-1,-0.4) to[out=0,in=180] (1,0.4);
			\draw[white,line width = 1.5mm] (-1,0.4) to[out=0,in=180] (1,-0.4);
			\draw[\myblue,line width = 0.5mm] (-1,0.4) to[out=0,in=180] (1,-0.4);
			\draw[\myblue,line width = 0.5mm] (-1,0.4) to[out=180,in=90] (-1.4,0) to[out=270,in=180] (-1,-0.4);
			\begin{scope}[xscale=-1]
				\draw[\myblue,line width = 0.5mm] (-1,0.4) to[out=180,in=90] (-1.4,0) to[out=270,in=180] (-1,-0.4);
			\end{scope}
		\end{tikzpicture}
	\end{array}=\begin{array}{c}
		\begin{tikzpicture}[scale=0.7]
			\draw[fill=white] (-1,0) circle (0.1) (1,0) circle (0.1);
			\draw[\myblue,line width = 0.5mm] (0,0.8) to[out=180,in=90] (-1.4,0) to[out=270,in=180] (-1,-0.4) to[out=0,in=180] (1,0.8) to[out=0,in=90] (1.8,0) to[out=270,in=0] (0,-0.8) to[out=180,in=270] (-1.8,0) to[out=90,in=180] (-1,0.8);
			\draw[white,line width = 1.5mm] (-1,0.8) to[out=0,in=180] (1,-0.4) to[out=0,in=270] (1.4,0) to[out=90,in=0] (0,0.8);
			\draw[\myblue,line width = 0.5mm] (-1,0.8) to[out=0,in=180] (1,-0.4) to[out=0,in=270] (1.4,0) to[out=90,in=0] (0,0.8);
		\end{tikzpicture}
	\end{array}\,.
	\label{mainid}
\end{equation}

Let us illustrate how the Kauffman bracket is applied on the l.h.s. of this expression:
\begin{equation}
	\begin{aligned}
		&\begin{array}{c}
			\begin{tikzpicture}[scale=0.7]
				\draw[fill=white] (-1,0) circle (0.1) (1,0) circle (0.1);
				\draw[\myblue,line width = 0.5mm] (-1,-0.4) to[out=0,in=180] (1,0.4);
				\draw[white,line width = 1.5mm] (-1,0.4) to[out=0,in=180] (1,-0.4);
				\draw[\myblue,line width = 0.5mm] (-1,0.4) to[out=0,in=180] (1,-0.4);
				\draw[\myblue,line width = 0.5mm] (-1,0.4) to[out=180,in=90] (-1.4,0) to[out=270,in=180] (-1,-0.4);
				\begin{scope}[xscale=-1]
					\draw[\myblue,line width = 0.5mm] (-1,0.4) to[out=180,in=90] (-1.4,0) to[out=270,in=180] (-1,-0.4);
				\end{scope}
			\end{tikzpicture}
		\end{array}=q^{\frac{1}{2}}\begin{array}{c}
			\begin{tikzpicture}[scale=0.7]
				\draw[fill=white] (-1,0) circle (0.1) (1,0) circle (0.1);
				\draw[\myblue,line width = 0.5mm] (-0.3,0) to[out=90,in=0] (-1,0.4) to[out=180,in=90] (-1.4,0) to[out=270,in=180] (-1,-0.4) to[out=0,in=270] (-0.3,0);
				\begin{scope}[xscale=-1]
					\draw[\myblue,line width = 0.5mm] (-0.3,0) to[out=90,in=0] (-1,0.4) to[out=180,in=90] (-1.4,0) to[out=270,in=180] (-1,-0.4) to[out=0,in=270] (-0.3,0);
				\end{scope}
			\end{tikzpicture}
		\end{array}+q^{-\frac{1}{2}}\begin{array}{c}
			\begin{tikzpicture}[scale=0.7]
				\draw[fill=white] (-1,0) circle (0.1) (1,0) circle (0.1);
				\draw[\myblue,line width = 0.5mm] (0,0.2) to[out=180,in=0] (-1,0.4) to[out=180,in=90] (-1.4,0) to[out=270,in=180] (-1,-0.4) to[out=0,in=180] (0,-0.2);
				\begin{scope}[xscale=-1]
					\draw[\myblue,line width = 0.5mm] (0,0.2) to[out=180,in=0] (-1,0.4) to[out=180,in=90] (-1.4,0) to[out=270,in=180] (-1,-0.4) to[out=0,in=180] (0,-0.2);
				\end{scope}
			\end{tikzpicture}
		\end{array}=q^{\frac{1}{2}}\begin{array}{c}
			\begin{tikzpicture}[scale=0.6]
				\begin{scope}[shift={(0,-0.25)}]
					\begin{scope}[yscale=0.8]
						\draw[\myblue,line width = 0.5mm] (-0.7,0) to[out=90,in=90] (-0.3,0);
						\draw[\myblue,line width = 0.5mm] (0.7,0) to[out=90,in=90] (0.3,0);
					\end{scope}
				\end{scope}
				\foreach \y in {0,1,2}
				{
					\begin{scope}[shift={(0,\y)}]
						\draw[line width = 1.5mm,white] (0.5,0) to[out=90,in=270] (-0.5,1);
						\draw[line width = 0.5mm,black] (0.5,0) to[out=90,in=270] (-0.5,1);
						\draw[line width = 1.5mm,white] (-0.5,0) to[out=90,in=270] (0.5,1);
						\draw[line width = 0.5mm,black] (-0.5,0) to[out=90,in=270] (0.5,1);
					\end{scope}
				}
				\draw[line width = 1.5mm,white] (-0.5,0) -- (-0.5,-0.5) (0.5,0) -- (0.5,-0.5);
				\draw[line width = 0.5mm,black] (-0.5,0) -- (-0.5,-0.5) (0.5,0) -- (0.5,-0.5);
				\draw[line width = 0.5mm,black] (-0.5,-0.5) to[out=270,in=270] (-1,-0.5) -- (-1,3) to[out=90,in=90] (-0.5,3);
				\draw[line width = 0.5mm,black] (0.5,-0.5) to[out=270,in=270] (1,-0.5) -- (1,3) to[out=90,in=90] (0.5,3);
				\begin{scope}[shift={(0,-0.25)}]
					\begin{scope}[yscale=0.8]
						\draw[line width = 1.5mm,white] (-0.7,0) to[out=270,in=270] (-0.3,0);
						\draw[\myblue,line width = 0.5mm] (-0.7,0) to[out=270,in=270] (-0.3,0);
						\draw[line width = 1.5mm,white] (0.7,0) to[out=270,in=270] (0.3,0);
						\draw[\myblue,line width = 0.5mm] (0.7,0) to[out=270,in=270] (0.3,0);
					\end{scope}
				\end{scope}
			\end{tikzpicture}
		\end{array}+q^{-\frac{1}{2}}\begin{array}{c}
			\begin{tikzpicture}[scale=0.6]
				\begin{scope}[shift={(0,-0.25)}]
					\begin{scope}[yscale=0.8]
						\draw[\myblue,line width = 0.5mm] (-0.7,0) to[out=90,in=90] (0.7,0);
					\end{scope}
				\end{scope}
				\foreach \y in {0,1,2}
				{
					\begin{scope}[shift={(0,\y)}]
						\draw[line width = 1.5mm,white] (0.5,0) to[out=90,in=270] (-0.5,1);
						\draw[line width = 0.5mm,black] (0.5,0) to[out=90,in=270] (-0.5,1);
						\draw[line width = 1.5mm,white] (-0.5,0) to[out=90,in=270] (0.5,1);
						\draw[line width = 0.5mm,black] (-0.5,0) to[out=90,in=270] (0.5,1);
					\end{scope}
				}
				\draw[line width = 1.5mm,white] (-0.5,0) -- (-0.5,-0.5) (0.5,0) -- (0.5,-0.5);
				\draw[line width = 0.5mm,black] (-0.5,0) -- (-0.5,-0.5) (0.5,0) -- (0.5,-0.5);
				\draw[line width = 0.5mm,black] (-0.5,-0.5) to[out=270,in=270] (-1,-0.5) -- (-1,3) to[out=90,in=90] (-0.5,3);
				\draw[line width = 0.5mm,black] (0.5,-0.5) to[out=270,in=270] (1,-0.5) -- (1,3) to[out=90,in=90] (0.5,3);
				\begin{scope}[shift={(0,-0.25)}]
					\begin{scope}[yscale=0.8]
						\draw[line width = 1.5mm,white] (-0.7,0) to[out=270,in=270] (0.7,0);
						\draw[\myblue,line width = 0.5mm] (-0.7,0) to[out=270,in=270] (0.7,0);
					\end{scope}
				\end{scope}
			\end{tikzpicture}
		\end{array}=\\
		&\overset{\eqref{operators},\eqref{Lp}}{=}q^{\frac{1}{2}}\mu^2\Psi_{3_1}(j)+q^{-\frac{1}{2}}L_1\Psi_{3_1}(j)\,.
	\end{aligned}
\end{equation}

Expanding both sides using the Kauffman bracket \eqref{Kauffman} we derive the following relation:
\begin{equation}\label{L2L1L0}
	q^{-\frac{1}{2}}\left(L_1+q\mu^2\right)\Psi_{3_1}(j)=\left(q^{-\frac{3}{2}}L_2+q^{-\frac{1}{2}}\mu^2L_1-q^{-\frac{3}{2}}(1+q^2-2q^2\mu^2)\right)\Psi_{3_1}(j)\,.
\end{equation}
This relation can be easily verified by applying \eqref{Jones}.

From this relation we see that link $L_2$ in a presence of the trefoil knot is decomposed in the basis mentioned above:
\begin{equation}\label{L2}
	L_2\Psi_{3_1}(j)=\left(q(1-\mu^2)L_1+1+q^2-q^2\mu^2\right)\Psi_{3_1}(j)\,.
\end{equation}

To acquire expansions for higher links $L_p$ we add $p$ rings to the lower part of the trefoil and consider resulting equivalences:
\begin{equation}\label{relat}
	\begin{array}{c}
		\begin{tikzpicture}
			\node(A) at (0,0) {$\begin{array}{c}
					\begin{tikzpicture}[scale=1]
						\draw[-stealth] (-1.5,-0.7) -- (-1.5,3.7);
						\node[left] at (-1.5,3.7) {$\scriptstyle t$};
						\draw[ultra thick] (-1.55,-0.25) -- (-1.45,-0.25) (-1.55,3) -- (-1.45,3);
						\node[left] at (-1.5,-0.25) {$0$};
						\node[left] at (-1.5,3) {$1$};
						\draw[\myblue,line width = 0.5mm] (1,3.5) to[out=180,in=90] (0.8,2) to[out=270,in=90] (0.3,1);
						\begin{scope}[shift={(0,-0.25)}]
							\begin{scope}[yscale=0.8]
								\draw[\myblue,line width = 0.5mm] (-0.7,0) to[out=90,in=90] (0.7,0);
							\end{scope}
						\end{scope}
						\foreach \y in {0,1,2}
						{
							\begin{scope}[shift={(0,\y)}]
								\draw[line width = 1.5mm,white] (0.5,0) to[out=90,in=270] (-0.5,1);
								\draw[line width = 0.5mm,black] (0.5,0) to[out=90,in=270] (-0.5,1);
								\draw[line width = 1.5mm,white] (-0.5,0) to[out=90,in=270] (0.5,1);
								\draw[line width = 0.5mm,black] (-0.5,0) to[out=90,in=270] (0.5,1);
							\end{scope}
						}
						\draw[line width = 1.5mm,white] (-0.5,0) -- (-0.5,-0.5) (0.5,0) -- (0.5,-0.5);
						\draw[line width = 0.5mm,black] (-0.5,0) -- (-0.5,-0.5) (0.5,0) -- (0.5,-0.5);
						\draw[line width = 0.5mm,black] (-0.5,-0.5) to[out=270,in=270] (-1,-0.5) -- (-1,3) to[out=90,in=90] (-0.5,3);
						\draw[line width = 1.5mm,white] (0.5,-0.5) to[out=270,in=270] (1.5,-0.5) -- (1.5,3) to[out=90,in=90] (0.5,3);
						\draw[line width = 0.5mm,black] (0.5,-0.5) to[out=270,in=270] (1.5,-0.5) -- (1.5,3) to[out=90,in=90] (0.5,3);
						\begin{scope}[shift={(0,-0.25)}]
							\begin{scope}[yscale=0.8]
								\draw[line width = 1.5mm,white] (-0.7,0) to[out=270,in=270] (0.7,0);
								\draw[\myblue,line width = 0.5mm] (-0.7,0) to[out=270,in=270] (0.7,0);
							\end{scope}
						\end{scope}
						\node[below] at (0,-0.7) {$\scriptstyle p$};
						\draw[line width = 1.5mm,white] (0.3,1) to[out=270,in=270] (1.2,2) to[out=90,in=0] (1,3.5);
						\draw[\myblue,line width = 0.5mm] (0.3,1) to[out=270,in=270] (1.2,2) to[out=90,in=0] (1,3.5);
					\end{tikzpicture}
				\end{array}$};
			\node(B) at (6,-1.3) {$\begin{array}{c}
					\begin{tikzpicture}[scale=0.7]
						\draw[fill=white] (-1,0) circle (0.1) (1,0) circle (0.1);
						\draw[\myblue,line width = 0.5mm] (0,0.8) to[out=180,in=90] (-1.4,0) to[out=270,in=180] (-1,-0.3) to[out=0,in=270] (-0.7,0) to[out=90,in=180] (0,0.5);
						\draw[\myblue,line width = 0.5mm] (-1,-0.8) to[out=0,in=180] (1,0.8);
						\begin{scope}[xscale=-1]
							\draw[white,line width = 1.5mm] (0,0.8) to[out=180,in=90] (-1.4,0) to[out=270,in=180] (-1,-0.3) to[out=0,in=270] (-0.7,0) to[out=90,in=180] (0,0.5);
							\draw[\myblue,line width = 0.5mm] (0,0.8) to[out=180,in=90] (-1.4,0) to[out=270,in=180] (-1,-0.3) to[out=0,in=270] (-0.7,0) to[out=90,in=180] (0,0.5);
						\end{scope}
						\draw[white,line width = 1.5mm] (-1,0.8) to[out=0,in=180] (1,-0.8);
						\draw[\myblue,line width = 0.5mm] (-1,0.8) to[out=0,in=180] (1,-0.8);
						\draw[\myblue,line width = 0.5mm] (-1,-0.8) to[out=180,in=270] (-1.6,0) to[out=90,in=180] (-1,0.8);
						\begin{scope}[xscale=-1]
							\draw[\myblue,line width = 0.5mm] (-1,-0.8) to[out=180,in=270] (-1.6,0) to[out=90,in=180] (-1,0.8);
						\end{scope}
						\node[above] at (0,0.8) {$\scriptstyle p$};
					\end{tikzpicture}
				\end{array}$};
			\node(C) at (6,1.3) {$\begin{array}{c}
					\begin{tikzpicture}[scale=0.7]
						\draw[fill=white] (-1,0) circle (0.1) (1,0) circle (0.1);
						\draw[\myblue,line width = 0.5mm] (0,0.8) to[out=180,in=90] (-1.4,0) to[out=270,in=180] (-1,-0.4) to[out=0,in=180] (1,0.8) to[out=0,in=90] (1.8,0) to[out=270,in=0] (0,-0.8) to[out=180,in=270] (-1.8,0) to[out=90,in=180] (-1,0.8);
						\draw[white,line width = 1.5mm] (-1,0.8) to[out=0,in=180] (1,-0.4) to[out=0,in=270] (1.4,0) to[out=90,in=0] (0,0.8);
						\draw[\myblue,line width = 0.5mm] (-1,0.8) to[out=0,in=180] (1,-0.4) to[out=0,in=270] (1.4,0) to[out=90,in=0] (0,0.8);
						\draw[\myblue,line width = 0.5mm] (0,1.2) to[out=180,in=90] (-2,0) to[out=270,in=180] (0,-1.2) to[out=0,in=270] (2,0) to[out=90,in=0] (0,1.2);
						\node[above] at (0,1.2) {$\scriptstyle p$};
					\end{tikzpicture}
				\end{array}$};
			\path (A) edge[->] node[above] {up} (C) (A) edge[->] node[below] {down} (B);
			\node[rotate=90] at (6,0) {=};
			\draw[-stealth] (0,2.7) -- (6,2.7) node[pos=0.5,above] {\scriptsize flattening};
			\node[right] at (7.5,1.3) {at $t=1$};
			\node[right] at (7.5,-1.3) {at $t=0$};
		\end{tikzpicture}
	\end{array}\,.
\end{equation}
For generic $p$ the upper part in \eqref{relat} looks simply as a composition of a link at $p=0$ with link $L_p$.
This operation just shifts all the indices of symbols by $p$ in the corresponding expansion:
\begin{equation}
	\begin{array}{c}
		\begin{tikzpicture}[scale=0.5]
			\draw[fill=white] (-1,0) circle (0.1) (1,0) circle (0.1);
			\draw[\myblue,line width = 0.5mm] (0,0.8) to[out=180,in=90] (-1.4,0) to[out=270,in=180] (-1,-0.4) to[out=0,in=180] (1,0.8) to[out=0,in=90] (1.8,0) to[out=270,in=0] (0,-0.8) to[out=180,in=270] (-1.8,0) to[out=90,in=180] (-1,0.8);
			\draw[white,line width = 1.5mm] (-1,0.8) to[out=0,in=180] (1,-0.4) to[out=0,in=270] (1.4,0) to[out=90,in=0] (0,0.8);
			\draw[\myblue,line width = 0.5mm] (-1,0.8) to[out=0,in=180] (1,-0.4) to[out=0,in=270] (1.4,0) to[out=90,in=0] (0,0.8);
			\draw[\myblue,line width = 0.5mm] (0,1.2) to[out=180,in=90] (-2,0) to[out=270,in=180] (0,-1.2) to[out=0,in=270] (2,0) to[out=90,in=0] (0,1.2);
			\node[above] at (0,1.2) {$\scriptstyle p$};
		\end{tikzpicture}
	\end{array}=\left(q^{-\frac{3}{2}}L_{p+2}+q^{-\frac{1}{2}}\mu^2L_{p+1}-q^{-\frac{3}{2}}(1+q^2-2q^2\mu^2)L_p\right)\Psi_{3_1}(j)\,.
\end{equation}
This relation might encourage someone to handle symbols $L_p$ as simple loop operators ``$L^p$'', however this action is incorrect.
If we mimic the ``multiplication'' by ``$L^p$'' of \eqref{L2L1L0}, i.e. if we  simply shift in turn all the indices in the l.h.s. and the r.h.s. of \eqref{L2L1L0}, the equality will be spoiled.

Applying Kauffman bracket \eqref{Kauffman}  to \eqref{L2L1L0} each time we derive a relation among $L_k\Psi_{3_1}(j)$ for $k=0,\ldots,p+2$, and using relations at the previous step we re-express all $L_k\Psi_{3_1}(j)$ in terms of $\Psi_{3_1}(j)$ and $L_1\Psi_{3_1}(j)$:
\begin{equation}\label{L3L4}
	\begin{aligned}
		& L_3\Psi_{3_1}(j) = \Big(\left(\mu ^4 q^2-3 \mu ^2 q^2+q^2+2\right)L_1+\left(q^5+\mu ^4 q^3-3 \mu ^2 q^3+q^3\right)\Big)\Psi_{3_1}(j)\,,\\
		& L_4\Psi_{3_1}(j) = \Big(\left(q^7-\mu ^6 q^3+5 \mu ^4 q^3-6 \mu ^2 q^3+q^3-3 \mu ^2 q+3 q\right)L_1+\\
		&\hspace{3cm}+\left(-\mu ^6q^4+5 \mu ^4 q^4-6 \mu ^2 q^4+q^4-3 \mu ^2 q^2+3 q^2+2\right)\Big)\Psi_{3_1}(j)\,,\\
		&\ldots
	\end{aligned}
\end{equation}

\subsection{Deriving A-polynomial}

By applying the flattening procedure to the cabling operator $\lambda$ we derive the following link:
\begin{equation}\label{lambda}
	\lambda\Psi_{3_1}(j)=\begin{array}{c}
		\begin{tikzpicture}[scale=0.8]
			\draw[fill=white] (-1,0) circle (0.1) (1,0) circle (0.1);
			\draw[\myblue,line width = 0.5mm] (-1.4,0) to[out=270,in=180] (-1,-0.4) to[out=0,in=180] (1,0.8) to[out=0,in=90] (1.8,-0.5);
			\draw[\myblue,line width = 0.5mm] (-1,-0.8) to[out=0,in=270] (0.5,0);
			\begin{scope}[yscale=-1]
				\draw[white,line width = 1.5mm] (-1.4,0) to[out=270,in=180] (-1,-0.4) to[out=0,in=180] (1,0.8) to[out=0,in=90] (1.8,-0.5);
				\draw[\myblue,line width = 0.5mm] (-1.4,0) to[out=270,in=180] (-1,-0.4) to[out=0,in=180] (1,0.8) to[out=0,in=90] (1.8,-0.5);
			\end{scope}
			\draw[\myblue,line width = 0.5mm] (1.8,0.5) to[out=90,in=0] (0,1) to[out=180,in=90] (-1.8,0) to[out=270,in=180] (-1,-0.8);
			\draw[white,line width = 1.5mm] (0.5,0) to[out=90,in=0] (-1,1) to[out=180,in=90] (-2.2,0) to[out=270,in=180] (0,-1) to[out=0,in=270] (1.8,-0.5);
			\draw[\myblue,line width = 0.5mm] (0.5,0) to[out=90,in=0] (-1,1) to[out=180,in=90] (-2.2,0) to[out=270,in=180] (0,-1) to[out=0,in=270] (1.8,-0.5);
		\end{tikzpicture}
	\end{array}\,.
\end{equation}
Applying the Kauffman bracket we find:
\begin{equation}
	\lambda\Psi_{3_1}(j)=\left(q^{-\frac{1}{2}}\mu L_2+q^{-\frac{3}{2}}\mu(1+q^2\mu^2)L_1+q^{\frac{3}{2}}\mu\left(-3+2\mu^2\right)\right)\Psi_{3_1}(j)\,.
\end{equation}

To derive the result of the action by $\lambda^2$ it suffices to cable the link in the r.h.s. of \eqref{lambda}.
The resulting expression turns out to be rather cumbersome.
So here we present expressions only after simplifying them with \eqref{L2}, \eqref{L3L4}:
\begin{equation}
	\begin{aligned}
		&\lambda\Psi_{3_1}(j)=\left(\frac{\mu  \left(q^2+1\right)}{q^{3/2}}L_1+\frac{\mu  \left(\left(\mu ^2-2\right) q^3+q\right)}{q^{3/2}}\right)\Psi_{3_1}(j)\,,\\
		&\lambda^2\Psi_{3_1}(j)=\Bigg(\frac{-\mu ^4+3 \mu ^2+\left(\mu ^4-3 \mu ^2+1\right) q^{10}-q^8+q^6-1}{q^5}L_1+\\
		&\hspace{2cm}+\frac{-\mu ^4+3 \mu ^2+\left(\mu ^6-5 \mu ^4+6 \mu ^2-1\right) q^{10}-q^8+q^6+2 q^4-1}{q^4}\Bigg)\Psi_{3_1}(j)\,.
	\end{aligned}
\end{equation}

Excluding a non-trivial link from the first expression and substituting into the second one we derive a quantum A-polynomial of operators $\lambda$ and $\mu$ annihilating $\Psi_{3_1}(j)$:
\begin{equation}
	\begin{aligned}
		&\CA_{3_1}(\lambda,\mu)=  q^{7/2} \lambda^2\left(q^2+1\right)\mu+(1-q^2)\lambda\big(\mu ^4-3 \mu ^2+\mu ^4 q^8-3 \mu ^2 q^8+\\
		&+q^8+\mu ^4 q^6-3 \mu ^2 q^6+\mu ^4 q^4-3 \mu ^2 q^4+q^4+\mu ^4 q^2-3 \mu ^2 q^2+q^2+1\big)+\\
		&+\mu  q^{3/2} \left(-\mu ^6+6 \mu ^4-10 \mu ^2+\mu ^2 q^{10}-q^{10}-\mu ^6 q^8+6 \mu ^4 q^8-10 \mu ^2 q^8+5 q^8+\mu ^2 q^6-3 q^6-2 q^4-2 q^2+3\right)\,.
	\end{aligned}
\end{equation}

By taking a quasi-classical limit $q\to 1$, $\lambda\to l+l^{-1}$, $\mu\to m+m^{-1}$ we derive a classical (overcomplicated) A-polynomial:
\begin{equation}
	A_{3_1}(l,m)=\frac{2 \left(m^2+1\right) \left(l-m^3\right) \left(l+m^3\right) \left(l m^3-1\right) \left(l m^3+1\right)}{l^2 m^7}\,.
\end{equation}

\subsection{Simplifying A-polynomial}

As we might have noted the A-polynomial derived in the previous section might be overcomplicated.
In particular, in the quasi-classical limit the A-polynomial factorizes and contains factors corresponding to the both left hand sided and right hand sided trefoil knots.
This effect might indicate that the choice of overdetermined linear system to construct a constraint was not optimal.

Since Wilson loop operators behave as $\sim\sinh x$, where $x$ is an actual classical coordinate on the moduli space of $SL(2,\IC)$ flat connections, there are no nice commutation relations on $\lambda$ and $\mu$, rather
\begin{equation}
	\mu=-(Q+Q^{-1}),\quad \lambda=P+P^{-1},\quad Q\,P=q^2\, P\,Q\,.
\end{equation}

As a variant of another overdetermined system let us consider $\Psi_{3_1}(j)$, $\lambda\Psi_{3_1}(j)$, $\mu\lambda\Psi_{3_1}(j)$.
For a new element we derive:
\begin{equation}
	\begin{aligned}
		& \mu\lambda\Psi_{3_1}(j)=\begin{array}{c}
			\begin{tikzpicture}[scale=0.8]
				\draw[fill=white] (-1,0) circle (0.1) (1,0) circle (0.1);
				\draw[\myblue,line width = 0.5mm] (-0.75,0) to[out=90,in=0] (-1,0.25) to[out=180,in=90] (-1.6,0);
				\draw[\myblue,line width = 0.5mm] (-1.4,0) to[out=270,in=180] (-1,-0.4) to[out=0,in=180] (1,0.8) to[out=0,in=90] (1.8,-0.5);
				\draw[\myblue,line width = 0.5mm] (-1,-0.8) to[out=0,in=270] (0.5,0);
				\begin{scope}[yscale=-1]
					\draw[white,line width = 1.5mm] (-1.4,0) to[out=270,in=180] (-1,-0.4) to[out=0,in=180] (1,0.8) to[out=0,in=90] (1.8,-0.5);
					\draw[\myblue,line width = 0.5mm] (-1.4,0) to[out=270,in=180] (-1,-0.4) to[out=0,in=180] (1,0.8) to[out=0,in=90] (1.8,-0.5);
				\end{scope}
				\draw[\myblue,line width = 0.5mm] (1.8,0.5) to[out=90,in=0] (0,1) to[out=180,in=90] (-1.8,0) to[out=270,in=180] (-1,-0.8);
				\draw[white,line width = 1.5mm] (0.5,0) to[out=90,in=0] (-1,1) to[out=180,in=90] (-2.2,0) to[out=270,in=180] (0,-1) to[out=0,in=270] (1.8,-0.5);
				\draw[\myblue,line width = 0.5mm] (0.5,0) to[out=90,in=0] (-1,1) to[out=180,in=90] (-2.2,0) to[out=270,in=180] (0,-1) to[out=0,in=270] (1.8,-0.5);
				\draw[white,line width = 1.5mm] (-0.75,0) to[out=270,in=0] (-1,-0.25) to[out=180,in=270] (-1.6,0);
				\draw[\myblue,line width = 0.5mm] (-0.75,0) to[out=270,in=0] (-1,-0.25) to[out=180,in=270] (-1.6,0);
			\end{tikzpicture}
		\end{array}=\\
		&=\left(\frac{\mu ^2+\mu ^2 q^4-q^4+2 q^2-1}{q^{5/2}}L_1+\frac{\mu ^2+\mu ^4 q^4-3 \mu ^2 q^4+q^4+\mu ^2 q^2-1}{q^{3/2}}\right)\Psi_{3_1}(j)\,.
	\end{aligned}
\end{equation}
Again, excluding $L_1\Psi_{3_1}(j)$ we derive a simpler ``square root'' A-polynomial:
\begin{equation}
	\CA^{\rm sqrt}_{3_1}(\lambda,\mu)=q \left(q^2+1\right)\mu\lambda\mu+\lambda\left(-\mu ^2-\mu ^2 q^4+q^4-2 q^2+1\right)-q^{3/2} (1-q^2)\mu \left(-\mu ^4+5 \mu ^2+q^2-5\right)\,.
	\label{rootpol}
\end{equation}

Or in terms of $P$ and $Q$:
\begin{equation}
	\CA^{\rm sqrt}_{3_1}(P,Q)=Q^3(q^2-Q^4)P+q^{3/2} \left(Q^2+1\right) \left(q^2 Q^4-Q^8+Q^6-Q^4+Q^2-1\right)+Q^3(q^2Q^4-1)P^{-1}\,.
\end{equation}
The quasi-classical limit of this quantum polynomial delivers the following expression:
\begin{equation}\label{A_sqrt}
	A_{3_1}^{\rm sqrt}(l,m)=(m^4-1)(l+m^3)(l m^3-1)\,.
\end{equation}

\subsection{Testing expressions}

Surely all the presented quantities could be calculated independently,
with the help of the Reshetikhin-Turaev (RT) formalism and its extension for cabling, see e.g. \cite{cabling}.   
For the trefoil we actually have:\footnote{Where we use the standard notation for the quantum number: $$[x]_q=\frac{q^x-q^{-x}}{q-q^{-1}}\,.$$}
\begin{equation}\label{Jones}
	\begin{aligned}
		&\Psi_{3_1}(j)=\sum\lm_{k=0}^{2j}(-1)^kq^{3(k(1-k)+4jk-2j^2)}\left[2(2j-k)+1\right]_q\,,\\
		&L_p\Psi_{3_1}(j)=\sum\lm_{k=0}^{2j}(-1)^kq^{3(k(1-k)+4jk-2j^2)}\left[2(2j-k)+1\right]_q\times\left(-q^{4j-2k+1}-q^{-4j+2k-1}\right)^p\,.
	\end{aligned}
\end{equation}
One could check easily that these expressions satisfy 
all the above relations --
from (\ref{L2L1L0}) to (\ref{rootpol}).

To avoid notational confusion let us present here our conventions for the quantum algebra $U_q(\fs\fu_2)$ implemented in the RT formalism throughout the paper:
\begin{equation}
	\left[h,e\right]=2e,\quad \left[h,f\right]=-2f,\quad \left[e,f\right]=\frac{q^h-q^{-h}}{q-q^{-1}}\,.
\end{equation}

A co-product structure for this algebra is defined as:
\begin{equation}
	\Delta(e)=e\otimes q^{\frac{h}{2}}+q^{-\frac{h}{2}}\otimes e,\quad \Delta(f)=f\otimes q^{\frac{h}{2}}+q^{-\frac{h}{2}}\otimes f,\quad \Delta(h)=h\otimes 1+1\otimes h\,.
\end{equation}

In these terms universal R-matrices are calculated as:
\begin{equation}
	\begin{split}
		&R=q^{-\frac{1}{4}h\otimes h}\left(\sum\lm_{k=0}^{\infty}\frac{(-1)^k\left(q-q^{-1}\right)^k}{[k]_q!}q^{-\frac{k(k-1)}{2}}\,f^k\otimes e^k\right) q^{-\frac{1}{4}h\otimes h}\,,\\
		&R^{-1}=q^{\frac{1}{4}h\otimes h}\left(\sum\lm_{k=0}^{\infty}\frac{\left(q-q^{-1}\right)^k}{[k]_q!}q^{\frac{k(k-1)}{2}}\,f^k\otimes e^k\right) q^{\frac{1}{4}h\otimes h}\,.
	\end{split}
\end{equation}

Eigenvalues of the R-matrices in a tensor product $j\otimes j$ in a representation of spin $2j-k$ read:
\begin{equation}
	r_{2j-k,j\otimes j}=(-1)^k q^{k(1-k)+4jk-2j^2}\,.
\end{equation}

Eventually, in a calculation one might need respective Clebsh-Gordan coefficients for vacuum transitions:
\begin{equation}
	\langle 0,0|m,j,m',j\rangle=\langle m,j,m',j|0,0\rangle=(-q)^m\delta_{m+m',0}\,.
\end{equation}

\section{Towards augmentation theory in quasi-classics}

In this section we sketch a quasi-classical transition to the augmentation theory and contact geometry,
which  deliver the (classical) augmentation polynomials.
It is a kind of important, still auxiliary story, which does not affect the main content of the paper.

\subsection{Clebsh-Gordan (CG) chords}

Clebsh-Gordan coefficients arise in the isotipical decomposition of quantum group representations:
\begin{equation}
	\begin{aligned}
		&|m_1,j_1\rangle\otimes|m_2,j_2\rangle=\sum\lm_{m_3,j_2}C_{m_1,j_1;m_2,j_2}^{m_3,j_3}|m_3,j_3\rangle\,,\\
		&|m_3,j_3\rangle=\sum\lm_{m_1,m_2}\bar C_{m_3,j_3}^{m_1,j_1;m_2,j_2}|m_1,j_1\rangle\otimes|m_2,j_2\rangle\,.
	\end{aligned}
\end{equation}

The Clebsh-Gordan elements can be used to extend the minimalist Reshetikhin-Turaev formalism consisting of 4-valent vertices representing R-matrices and marked arcs representing vacuum transitions to incorporate 3-valent vertices \cite{Moore:1988qv}.
These 3-valent vertices represent choices in the isotipical components:
\begin{equation}
	C_{\frac{1}{2};j}^{j\pm\frac{1}{2}}\to \begin{array}{c}
		\begin{tikzpicture}
			\draw[thick] (0.3,-0.3) -- (0,0) -- (0,0.5);
			\draw[\myblue] (-0.3,-0.3) -- (0,0);
			\node[below] at (0.3,-0.3) {$\scriptstyle j$};
			\node[above] at (0,0.5) {$\scriptstyle j\pm\frac{1}{2}$};
		\end{tikzpicture}
	\end{array},\quad C_{j;\frac{1}{2}}^{j\pm\frac{1}{2}}\to \begin{array}{c}
		\begin{tikzpicture}[xscale=-1]
			\draw[thick] (0.3,-0.3) -- (0,0) -- (0,0.5);
			\draw[\myblue] (-0.3,-0.3) -- (0,0);
			\node[below] at (0.3,-0.3) {$\scriptstyle j$};
			\node[above] at (0,0.5) {$\scriptstyle j\pm\frac{1}{2}$};
		\end{tikzpicture}
	\end{array},\quad \bar C^{\frac{1}{2};j}_{j\pm\frac{1}{2}}\to \begin{array}{c}
		\begin{tikzpicture}[yscale=-1]
			\draw[thick] (0.3,-0.3) -- (0,0) -- (0,0.5);
			\draw[\myblue] (-0.3,-0.3) -- (0,0);
			\node[above] at (0.3,-0.3) {$\scriptstyle j$};
			\node[below] at (0,0.5) {$\scriptstyle j\pm\frac{1}{2}$};
		\end{tikzpicture}
	\end{array},\quad \bar C^{j;\frac{1}{2}}_{j\pm\frac{1}{2}}\to \begin{array}{c}
		\begin{tikzpicture}[scale=-1]
			\draw[thick] (0.3,-0.3) -- (0,0) -- (0,0.5);
			\draw[\myblue] (-0.3,-0.3) -- (0,0);
			\node[above] at (0.3,-0.3) {$\scriptstyle j$};
			\node[below] at (0,0.5) {$\scriptstyle j\pm\frac{1}{2}$};
		\end{tikzpicture}
	\end{array}
\end{equation}

Using these new vertices we could describe skein relations for  higher  representations.
For example,
\begin{equation}\label{skein}
	\begin{aligned}
		&q^{-j-1}\begin{array}{c}
			\begin{tikzpicture}
				\draw[\myblue] (-0.5,-0.5) -- (0.5,0.5);
				\draw[thick] (0.5,-0.5) -- (0.05,-0.05) (-0.05,0.05) -- (-0.5,0.5); 
				\node[below] at (0.5,-0.5) {$\scriptstyle j$};
				\node[above] at (-0.5,0.5) {$\scriptstyle j$};
			\end{tikzpicture}
		\end{array}-q^{j+1}\begin{array}{c}
			\begin{tikzpicture}
				\draw[\myblue] (-0.5,-0.5) -- (-0.05,-0.05) (0.05,0.05) --  (0.5,0.5);
				\draw[thick] (0.5,-0.5) -- (-0.5,0.5); 
				\node[below] at (0.5,-0.5) {$\scriptstyle j$};
				\node[above] at (-0.5,0.5) {$\scriptstyle j$};
			\end{tikzpicture}
		\end{array}=\left(q^{-2j-1}-q^{2j+1}\right)\begin{array}{c}
			\begin{tikzpicture}
				\draw[thick] (0.5,-0.5) -- (0,0) -- (0,0.3) -- (-0.5,0.8);
				\draw[\myblue] (-0.5,-0.5) -- (0,0) (0,0.3) -- (0.5,0.8);
				\node[below] at (0.5,-0.5) {$\scriptstyle j$};
				\node[above] at (-0.5,0.8) {$\scriptstyle j$};
				\node[right] at (0,0.15) {$\scriptstyle j+\frac{1}{2}$};
			\end{tikzpicture}
		\end{array}\,\\
		&q^{j+1}\begin{array}{c}
			\begin{tikzpicture}[xscale=-1]
				\draw[\myblue] (-0.5,-0.5) -- (0.5,0.5);
				\draw[thick] (0.5,-0.5) -- (0.05,-0.05) (-0.05,0.05) -- (-0.5,0.5); 
				\node[below] at (0.5,-0.5) {$\scriptstyle j$};
				\node[above] at (-0.5,0.5) {$\scriptstyle j$};
			\end{tikzpicture}
		\end{array}-q^{-j-1}\begin{array}{c}
			\begin{tikzpicture}[xscale=-1]
				\draw[\myblue] (-0.5,-0.5) -- (-0.05,-0.05) (0.05,0.05) --  (0.5,0.5);
				\draw[thick] (0.5,-0.5) -- (-0.5,0.5); 
				\node[below] at (0.5,-0.5) {$\scriptstyle j$};
				\node[above] at (-0.5,0.5) {$\scriptstyle j$};
			\end{tikzpicture}
		\end{array}=\left(q^{2j+1}-q^{-2j-1}\right)\begin{array}{c}
			\begin{tikzpicture}[xscale=-1]
				\draw[thick] (0.5,-0.5) -- (0,0) -- (0,0.3) -- (-0.5,0.8);
				\draw[\myblue] (-0.5,-0.5) -- (0,0) (0,0.3) -- (0.5,0.8);
				\node[below] at (0.5,-0.5) {$\scriptstyle j$};
				\node[above] at (-0.5,0.8) {$\scriptstyle j$};
				\node[right] at (0,0.15) {$\scriptstyle j+\frac{1}{2}$};
			\end{tikzpicture}
		\end{array}\,.
	\end{aligned}
\end{equation}

Considering WZW conformal blocks as CS wave-functions at fixed temporal slices of $S^3$ we think of braids as certain evolution operators acting in the respective Hilbert space.
In addition to standard evolution operators consisting of the R-matrices we could introduce new \emph{Clebsh-Gordan (CG) chord} operators in the following way.
First we ``peel off'' a spin-$\sfrac{1}{2}$ strand from the thick spin-$j$ cable, then we parallel transport it using Knizhnik-Zamolodchikov connection, and, finally, ``glue it to'' some other thick strand.
Diagrammatically these new chords are easily visualizable via 3-valent Clebsh-Gordan vertices:
\begin{equation}
	\Theta_{a,b}:=\left(1-h_b^4\right)\times\left\{\begin{array}{ll}
		\begin{array}{c}
			\begin{tikzpicture}[scale=0.5]
				\foreach \i in {-2,-1,0,4,5,6}
				{
					\draw[thick] (\i,-0.5) -- (\i,1);
				}
				\foreach \i in {1,2,3}
				{
					\draw[thick] (\i,-0.5) -- (\i,0.125*\i-0.2);
					\draw[thick] (\i,0.125*\i+0.2) -- (\i,1);
				}
				\draw[\myblue] (0,0) -- (4,0.5);
				\node[below] at (0,-0.5) {$\scriptstyle a$};
				\node[below] at (4,-0.5) {$\scriptstyle b$};
			\end{tikzpicture}
		\end{array}, & a<b\,;\\
		\begin{array}{c}
			\begin{tikzpicture}[scale=0.5,xscale=-1]
				\foreach \i in {-2,-1,0,4,5,6}
				{
					\draw[thick] (\i,-0.5) -- (\i,1);
				}
				\foreach \i in {1,2,3}
				{
					\draw[thick] (\i,-0.5) -- (\i,0.125*\i-0.2);
					\draw[thick] (\i,0.125*\i+0.2) -- (\i,1);
				}
				\draw[\myblue] (0,0) -- (4,0.5);
				\node[below] at (0,-0.5) {$\scriptstyle a$};
				\node[below] at (4,-0.5) {$\scriptstyle b$};
			\end{tikzpicture}
		\end{array}, & a>b\,;
	\end{array}\right.
\end{equation}
where $h=q^{j}$ with $j$ corresponding to the spin of the representation of the respective strand.

We could not help emphasizing apparent similarities between link symbols $L_p$ in \eqref{Lp} and CG chords.
Both are (a combination of) matrix coefficients of parallel transport -- holonomy -- for the flat Knizhnik-Zamolodchikov connection along paths defined by links.
However the CG chords correspond to single matrix elements and open paths, whereas the link symbols correspond to quantum traces of closed loops.
We would not pursue an exact relation in this letter, however let us note that it would be much simpler to derive a relation in the quasi-classical limit where WKB techniques are applicable to the Knizhnik-Zamolodchikov equation (see e.g. \cite{Dimofte:2009yn,Gaiotto:2011nm,Terashima:2013fg,Galakhov:2014aha,Galakhov:2015gza} and \cite{Neitzke:2021gxr,Freed:2022yae} for more recent discussions).

\subsection{Braid action on CG chords}

It is much simpler to make the transport of spin-$\sfrac{1}{2}$ strands through intersections we used in the flattening procedure in \eqref{braid} algorithmic in terms of CG chords.

The braid group generators $\sigma_k$ twisting the $k^{\rm th}$ and $(k+1)^{\rm th}$ strands act on the CG chords by conjugations.
For example,
\begin{equation}
	\sigma_k\left[\begin{array}{c}
		\begin{tikzpicture}[yscale=0.7]
			\draw[thick] (0,0) -- (0,1) (1,0) -- (1,1);
			\draw[\myblue] (0,0.25) -- (1,0.75);
			\node[below] at (0,0) {$\scriptstyle k$};
			\node[below] at (1,0) {$\scriptstyle k+1$};
		\end{tikzpicture}
	\end{array}\right]=\begin{array}{c}
		\begin{tikzpicture}[yscale=0.7]
			\draw[thick] (0,0) -- (0,1) (1,0) -- (1,1);
			\draw[\myblue] (0,0.25) -- (1,0.75);
			\node[below] at (0,-0.5) {$\scriptstyle k$};
			\node[below] at (1,-0.5) {$\scriptstyle k+1$};
			\begin{scope}[shift={(0,1)}]
				\draw[thick] (1,0) to[out=90,in=270] (0,0.5);
				\draw[line width = 1mm, white] (0,0) to[out=90,in=270] (1,0.5);
				\draw[thick] (0,0) to[out=90,in=270] (1,0.5);
			\end{scope}
			\begin{scope}[yscale=-1]
				\draw[thick] (1,0) to[out=90,in=270] (0,0.5);
				\draw[line width = 1mm, white] (0,0) to[out=90,in=270] (1,0.5);
				\draw[thick] (0,0) to[out=90,in=270] (1,0.5);
			\end{scope}
		\end{tikzpicture}
	\end{array}=\begin{array}{c}
		\begin{tikzpicture}[yscale=0.7]
			\draw[thick] (0,0) --(0,0.4) (0,0.6) -- (0,1) (1,0) -- (1,1);
			\draw[\myblue] (0,0.75) to[out=180,in=90] (-0.2,0.625) to[out=270,in=180] (0,0.5) -- (0.95,0.5) (1.05,0.5) to[out=0,in=90] (1.2,0.375) to[out=270,in=0] (1,0.25);
			\node[below] at (0,0) {$\scriptstyle k$};
			\node[below] at (1,0) {$\scriptstyle k+1$};
		\end{tikzpicture}
	\end{array}\,.
\end{equation}
Translating these graphs to operators we derive: $\sigma_{k}[\Theta_{k,k+1}]=h_kh_{k+1}^{-1}\Theta_{k+1,k}$.

If during the braiding spin-$\sfrac{1}{2}$ strands are linked with spin-$j$ strands we resolve these links by applying generalized skein relations \eqref{skein}.
This always allows one to re-expand linked CG chords as polynomials od unlinked CG chords.
As a result of these calculations we derive the following representation of the braid group on CG chords:
\begingroup
\renewcommand*{\arraystretch}{1.3}
\begin{equation}\label{chord_braid}
	\sigma_k:\left\{\begin{array}{c}
		h_k\mapsto h_{k+1},\quad h_{k+1}\mapsto h_{k},\quad h_i\mapsto h_i,\;i\neq k,k+1,\quad \Theta_{i,j}\mapsto\Theta_{i,j},\;i,j\neq k,k+1\,,\\
		\Theta_{k,k+1}\mapsto h_kh_{k+1}^{-1}\Theta_{k+1,k},\quad \Theta_{k+1,k}\mapsto h_k^{-1}h_{k+1}\Theta_{k,k+1}\,,\\
		\Theta_{i,k}\mapsto\Theta_{i,k+1},\;i\neq k,k+1,\quad \Theta_{k,j}\mapsto\Theta_{k+1,j},\;j\neq k,k+1\;,\\
		\Theta_{i,k+1}\mapsto h_{k+1}^2\Theta_{i,k}+ h_k h_{k+1}^{-2}\Theta_{i,k+1}\Theta_{k+1,k},\;i<k\,,\\
		\Theta_{i,k+1}\mapsto h_{k+1}^2\Theta_{i,k}+h_{k+1}^{-1}\Theta_{i,k+1}\Theta_{k+1,k},\;i>k+1\,,\\
		\Theta_{k+1,j}\mapsto h_{k+1}^{-2}\Theta_{k,j}- h_{k+1}^{-3}\Theta_{k,k+1}\Theta_{k+1,j},\;j>k+1\,,\\
		\Theta_{k+1,j}\mapsto   h_{k+1}^{-2}\Theta_{k,j}- h_k^{-1}h_{k+1}^{-2}\Theta_{k,k+1}\Theta_{k+1,j},\;j<k\,.
	\end{array}\right\}
\end{equation}
\endgroup

It is easy to check that $\sigma_k$ generate indeed the braid group:
\begin{equation}
	\begin{aligned}
		&\sigma_k\circ\sigma_k^{-1}=\bbone,\quad\sigma_i\circ\sigma_j=\sigma_j\circ\sigma_i,\quad \mbox{for } |i-j|>2\,,\\
		& \sigma_{k}\circ\sigma_{k+1}\circ\sigma_{k}=\sigma_{k+1}\circ\sigma_{k}\circ\sigma_{k+1}\,.
	\end{aligned}
\end{equation}

\subsection{Reeb chords and augmentation polynomials}

Augmentation theory arises naturally in the quasi-classical limit.
In this limit we tend spin-$j$ coloring of the knot to infinity and $q\to 1$ so that $h=q^j$ remains constant.
Since the spin shift by $\pm\sfrac{1}{2}$ is small compared to $j$ the CG chords may be considered in the presence of braid closures -- links -- moreover they acquire vacuum expectation values corresponding to parallel transports with respect to an $SL(2,\IC)$-connection in the knot complement in one of CS path integral saddle points:
\begin{equation}
	q\to1,\quad q^j\to m,\quad \Theta_{a,b}\longrightarrow\theta_{a,b}\,.
\end{equation} 
Coordinates $\theta_{a,b}$ are coordinates on the quasi-classical $SL(2,\IC)$ moduli space and resemble \emph{Reeb chords} saturating the knot contact homology groups in degree 0 (for definitions see \cite{1210.4803}).
Actually, the quasi-classical CG chords are images of the augmentation map applied to the Reeb chords in degree 0.

We would like to argument the latter statement in two steps.
First of all let us note that the quasi-classical limit of the braid group action \eqref{chord_braid} is equivalent to the braid group action on the Reeb chords \cite[definition 3.3]{1210.4803}.
Furthermore we apply the same strategy to derive augmentation polynomials (classical A-polynomials).

As we acted in the case of \eqref{two_paths} consider two paths of transforming a CG chord on a knot that is represented as a braid $\CB$ closure: one end of the chord we braid with $\CB$,  the other end is dragged to the same time slice via the braid closure:
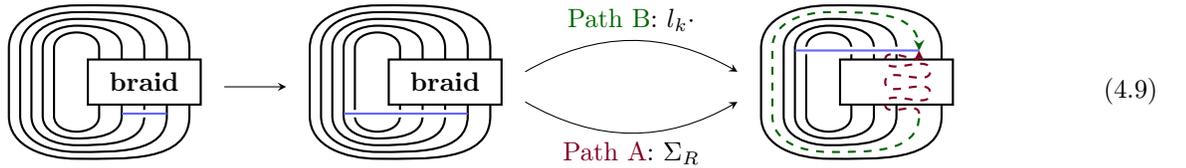
\begin{equation}
	\begin{array}{c}
		\begin{tikzpicture}
			\node (A) at (0,0) {$\begin{array}{c}
					\begin{tikzpicture}[scale=0.3]
						\draw[thick] (-2.5,-1) -- (-2.5,1) -- (2.5,1) -- (2.5,-1) -- cycle;
						\foreach \n in {0,1,2,3,4}
						{
							\draw[thick] (\n-2, 1) -- (\n-2,1.5) to[out=90,in=0] (-3,2.2+0.3*\n) to[out=180,in=90] (-4-0.5*\n,1.5) -- (-4-0.5*\n,-1.5) to[out=270,in=180] (-3,-2.2-0.3*\n) to[out=0,in=270] (\n-2,-1.5) -- (\n-2, -1); 
						}
						\node at (0,0) {\bf braid};
						\draw[white, line width = 1mm] (-0.5,-1.4) -- (0.5,-1.4); 
						\draw[\myblue, thick] (-1,-1.4) -- (1,-1.4);
					\end{tikzpicture}
				\end{array}$};
			\node (B) at (4,0) {$\begin{array}{c}
					\begin{tikzpicture}[scale=0.3]
						\draw[thick] (-2.5,-1) -- (-2.5,1) -- (2.5,1) -- (2.5,-1) -- cycle;
						\foreach \n in {0,1,2,3,4}
						{
							\draw[thick] (\n-2, 1) -- (\n-2,1.5) to[out=90,in=0] (-3,2.2+0.3*\n) to[out=180,in=90] (-4-0.5*\n,1.5) -- (-4-0.5*\n,-1.5) to[out=270,in=180] (-3,-2.2-0.3*\n) to[out=0,in=270] (\n-2,-1.5) -- (\n-2, -1); 
						}
						\node at (0,0) {\bf braid};
						\draw[white, line width = 1mm] (-4.2,-1.4) -- (0.5,-1.4); 
						\draw[\myblue, thick] (-4.5,-1.4) -- (1,-1.4);
					\end{tikzpicture}
				\end{array}$};
			\node (C) at (10,0) {$\begin{array}{c}
					\begin{tikzpicture}[scale=0.3]
						\draw[thick] (-2.5,-1) -- (-2.5,1) -- (2.5,1) -- (2.5,-1) -- cycle;
						\foreach \n in {0,1,2,4}
						{
							\draw[thick] (\n-2, 1) -- (\n-2,1.5) to[out=90,in=0] (-3,2.2+0.3*\n) to[out=180,in=90] (-4-0.5*\n,1.5) -- (-4-0.5*\n,-1.5) to[out=270,in=180] (-3,-2.2-0.3*\n) to[out=0,in=270] (\n-2,-1.5) -- (\n-2, -1); 
						}
						\draw[white, line width = 1mm] (-4.2,1.4) -- (0.5,1.4); 
						\draw[\myblue, thick] (-4.5,1.4) -- (1,1.4);
						\draw[thick, dashed, -stealth, burgundy] (1,-1.4) to[out=90,in=270] (-0.5,-1) to[out=90,in=270] (1.5, -0.5) to[out=90,in=270] (-0.5, 0) to[out=90,in=270] (1.5, 0.5) to[out=90,in=270] (-0.5, 1) to[out=90,in=270]   (1,1.4);
						\foreach \n in {3}
						{
							\draw[thick, dashed, stealth-, black!60!green] (\n-2, 1.4) -- (\n-2,1.5) to[out=90,in=0] (-3,2.2+0.3*\n) to[out=180,in=90] (-4-0.5*\n,1.5) -- (-4-0.5*\n,-1.5) to[out=270,in=180] (-3,-2.2-0.3*\n) to[out=0,in=270] (\n-2,-1.5) -- (\n-2, -1.4); 
						}
					\end{tikzpicture}
				\end{array}$};
			\draw[-stealth] (A.east) -- (B.west);
			\draw[-stealth] ([shift={(0,0.2)}]B.east) to[out=30,in=150] node[pos=0.5,above] {{\color{black!60!green} Path B}: $l_k\cdot$} ([shift={(0,0.2)}]C.west);
			\draw[-stealth] ([shift={(0,-0.2)}]B.east) to[out=-30,in=210] node[pos=0.5,below] {{\color{burgundy} Path A}: $\Sigma_R$} ([shift={(0,-0.2)}]C.west);
		\end{tikzpicture}
	\end{array}
\end{equation}

In the case A we think of variable $\theta_{a,b}$ as a variable $\theta_{0,b}$ where the left end has the index lower than all the other strand indices in the braid.
Then we apply sequence $\Sigma_R$ of morphisms $\sigma_k$ \eqref{chord_braid} as the respective braid generators appear in $\CB$.

In the case B we peel off spin-$\sfrac{1}{2}$ from strand $k$.
This is an analog of cabling operation we denote here as $l_k$.
To shift the spin of the whole knot we have to peel all the cables, so the quasi-classical shift operator expectation value reads:
\begin{equation}
	l=\prod\lm_k l_k\,.
\end{equation}

The augmentation polynomial arises if we exclude all the variables $\theta_{a,b}$ from the relations for the chords equating paths A and B:
\begin{equation}\label{aug}
	l_a \theta_{a,b}=\Sigma_R(\theta_{a,b})\,.
\end{equation}
In this way we have arrived to the image of the Reeb chords in degree 1 for the contact homology differential \cite[definition 3.11(2)]{1210.4803}.

\subsection{Trefoil again}

To illustrate the work of this mechanism let us consider again the simplest case of the trefoil knot.
In this case the knot is represented as a braid $\sigma_1^{\circ 3}$. 
We have 4 chords: $\theta_{1,1}$, $\theta_{1,2}$, $\theta_{2,1}$, $\theta_{2,2}$.

In this case we arrive to the following relations among chords:\footnote{In this derivation we adopt the construction of \cite{1210.4803} to match the augmentation polynomial literally. 
	Apparently, comparing our notations with ones in the above work we would note that certain change of coordinates, rescaling etc., are involved.}
\begin{equation}\label{aug_tre}
	\begin{aligned}
	&l_1 \left(m+1\right)+\theta_{1,2} \left(m \left(\theta_{1,2} \theta_{2,1}-2\right)-1\right)=0\,,\\
	&l_2 \theta_{1,2}+m \left(1-\theta_{1,2} \theta_{2,1}\right)+1=0 \,,\\
	&-\theta_{1,2} \theta_{2,1}+m \left(\theta_{1,2}^2 \theta_{2,1}^2+l_1 \theta_{2,1}-3 \theta_{1,2} \theta_{2,1}+1\right)+1=0\,,\\
	&l_2 \left(m+1\right)+\theta_{2,1} \left(m \left(2-\theta_{1,2} \theta_{2,1}\right)+1\right)=0\,.
	\end{aligned}
\end{equation}
Let us solve the second equation in \eqref{aug_tre} for $\theta_{1,2}=(m+1)(m \theta_{2,1}-l_2)^{-1}$.
After substitution the fourth equation reads:
\begin{equation}
	\frac{\left(\theta_{2,1}+l_2\right) \left(m^2 \left(-\theta_{2,1}\right)+l_2 m+l_2\right)}{l_2-m \theta_{2,1}}=0\,.
\end{equation}
This equation has two apparent roots.
Substituting either of roots we derive the following reductions of the first and the third equations in \eqref{aug_tre}:
\begin{equation}
	\begin{aligned}
		& \theta_{2,1}=-l_2,\quad  & \frac{\left(l_1 l_2+1\right) \left(m+1\right)}{l_2} =0,\quad -\left(l_1 l_2+1\right) m =0\,,\\
		& \theta_{2,1}=l_2\frac{m+1}{m^2},\quad & \frac{\left(m+1\right) \left(m^3+l_1 l_2\right)}{l_2}=0,\quad \frac{\left(m+1\right) \left(m^3+l_1 l_2\right)}{m}=0\,. 
	\end{aligned}
\end{equation}
Substituting $l_1l_2=l$ we arrive to the augmentation polynomial for the trefoil knot capturing all the above roots:
\begin{equation}\label{aug_3_1}
	{\rm Aug}_{3_1}(l,m)=(l+1)(m+1)(l+m^3)\,.
\end{equation}
Both polynomials \eqref{aug_3_1} and \eqref{A_sqrt} have a common root $l=-m^3$.

\section{Conclusion}

In this note we explained a simple way to calculate the quantum ${\cal A}$-polynomials
and their quasi-classical limits -- reproducing the ordinary $A$-polynomials.
The way is simple both conceptually and technically -- in fact, much simpler than the
quasi-classical study of Stokes lines and augmentation varieties,
what should make the subject available to a broader audience.
For simplicity we concentrated on the case of symmetrically-colored Jones polynomial
for the trefoil.

\bigskip

The obvious generalizations are hierarchical, because of the variety of questions,
which need to be solved.

\begin{itemize}
	
	\item{}
	The simplest one should be generalization to Jones polynomials of other knots --
	from arbitrary $2$-strand to arborescent \cite{Mironov:2015aia,Mironov:2015qma} and, further, 
	to arbitrary $m$-strand ones.
	For the $2$-strand case  just the complexity of the link diagrams 
	in (\ref{mainid}) and (\ref{lambda}) increases -- 
	it is easy to handle this with some space imagination.
	
	\item{}
	A little more involved problem is the case of Jones for more complicated knots --
	here one can think about developing some more algebraic methods instead of just drawing.
	This is also straightforward.
	
	\item{} 
	The Jones case is distinguished as compared to a theory with a more generic gauge group $SU(N)$ with $N>2$ by applicability of 
	Kauffman planarization, which provide considerable simplification in reducing
	sophisticated link diagrams to simple bases.
	It is an interesting question, if the generalization of this planarization to
	arbitrary $N$ for the special (actually, enormously large) set of bipartite knots
	\cite{Anokhina:2024lbn,ALM} can be applied to this problem.
	
	\item{}
	Generalization to arbitrary $N$ is especially interesting, because here the non-symmetric
	representations come into the game, and the entire set of colored polynomials
	and Ward identities become much richer and more interesting.
	This is, however, a rather complicated problem -- 
	and it requires going beyond Kauffman calculus.  
	
	\item{}
	At the same time for arbitrary $N$ a new powerful approach is provided by differential
	expansions \cite{Itoyama:2012fq,Mironov:2013qaa,Bishler:2020kqw}, which switches attention from ${\cal A}$- to ${\cal C}$-polynomials \cite{GarC,2009.11641},
	which can be more fundamental, because they are closer to the true content of the moduli space 
	(roughly speaking, better describe the variety of all knots).
	It is an interesting task to efficiently adapt our method so that it provides the 
	${\cal C}$-polynomials directly, 
	avoiding the intermediate step with ${\cal A}$-polynomials.
	
	\item{}
	The passing process from quantum operators and symbols in the knot complement incorporates a quasi-classical approximation, whose role in this consideration seems rather overrated.
	Indeed it is much simpler to work with non-commutative coordinates rather than quantum operators, yet the construction of the braid group action \eqref{chord_braid} on CG chords is generic in the full quantum theory.
	So there is a natural question if the augmentation theory might be intrinsically refined to capture the quantum deformations.
	Moreover since the resulting structure is a close cousin of the link algebra on a Riemann surface such a refinement might be extended to the second $t$-deformation of \cite{Aganagic:2011sg,Arthamonov:2015rha,Arthamonov:2017oxw}.

\end{itemize}

We hope that the work on these issues would deliver a new momentum with development
and a conceptual clarity of the simple enough technique, presented in this note.

\section*{Acknowledgments}
We would like to thank E.~Lanina, A.~Mironov and A.~Popolitov for illuminating discussions.
The work was partially funded within the state assignment of the Institute for Information Transmission Problems of RAS.
Our work is partly supported by grant RFBR 21-51-46010 ST\_a (D.G., A.M.), by the grants of the Foundation for the Advancement
of Theoretical Physics and Mathematics ``BASIS'' (A.M.).


\bibliographystyle{utphys}
\bibliography{biblio}

\end{document}